\documentclass[useAMS,usenatbib]{mn2e}

\usepackage{epsfig}

\title[Selection Issues in $z\ge5$ Surveys]{On Contamination and
  Completeness in $\mathbf{z\ge5}$ Lyman Break Galaxy Surveys}
\author[E R Stanway et al.]{Elizabeth R.~Stanway$^{1}$, Malcolm N. Bremer$^{1}$, Matthew D. Lehnert$^{2}$,
\\ $^1$H H Wills Physics Laboratory, Tyndall Avenue, Bristol, BS8 1TL, UK
\\$^2$Laboratoire d'Etudes des Galaxies, Etoiles, Physique et Instrumentation GEPI, Observatoire de Paris, Meudon, France}

\begin{document}

\date{Accepted . Received ; in original form }

\pagerange{\pageref{firstpage}--\pageref{lastpage}} \pubyear{}

\maketitle

\label{firstpage}

\begin{abstract}
  A large population of $z>5$ Lyman break galaxies has been identified
  in recent years.  However, the high redshift galaxies selected by
  different surveys are subject to a variety of selection effects -
  some overt, others more subtle.  We present an analysis of sample
  completeness and contamination issues in high redshift surveys,
  focusing on surveys at $z\approx5$ and using a
  spectroscopically-confirmed low redshift sample from the DEEP2
  survey in order to characterise contaminant galaxies. We find that
  most surveys underestimate their contamination from highly clustered
  galaxies at $z\approx1$ and stars. We consider the consequences of
  this for both the rest-frame ultraviolet luminosity function and the
  clustering signal from $z\approx5$ galaxies. We also find that
  sources with moderate strength Lyman-$\alpha$ emission lines can be
  omitted from dropout surveys due to their blue colours, again
  effecting the derived luminosity functions.  We discuss the points
  of comparison between different samples, and the applicability of
  survey-specific results to the population at $z>5$ in general.
\end{abstract}

\section{Introduction}
\label{sec:intro}

The study of $z>5$ galaxies has become a well-developed field in recent
years. The Lyman Break (or `dropout') technique - which identifies
starbursting sources by the dramatic spectral break around the
rest-frame Lyman-alpha feature and was first applied at $z\approx3$
\citep{1999ApJ...519....1S} - is now widely applied on large
multicolour datasets. As a result, large samples of
photometrically-selected high-redshift candidates can be constructed
with a comparatively small investment of telescope time.
Such photometric samples have been widely used to derive statistical
properties for $z>5$ galaxies as a whole, including luminosity
functions and clustering parameters.
\citep{2004ApJ...611..685O,2007MNRAS.376.1557I,2006ApJ...642...63L}

However, the high redshift galaxies selected by different surveys are
subject to a variety of selection effects - some overt, others more
subtle. While samples of Lyman-$\alpha$ emitting galaxies (selected
over a narrow redshift range, and relatively easy to confirm
spectroscopically due to the presence of strong emission lines) have
comparatively simple selection windows \citep[see][for
discussion]{2006A&A...460..681H}, the variety in width and shape of
broadband filters leads to a wide range of redshifts (and hence
intrinsic luminosity limits) falling into the selection window, while
the presence or absence of emission lines can alter the effectiveness
of a colour selection. Such effects are highly sensitive to filter
profile, selection criteria and survey depth.  Although attempts to
interpret the results of photometric surveys have generally considered
each such survey comparable, these sample-specific issues relating to
completeness and contamination can alter the resulting physical
interpretations. The situation is complicated by the fact that the
dominant contaminant populations have not hitherto been carefully and
systematically characterised.


While most surveys take into account a subset of these effects, the
description of published data is in many cases inadequate to allow
comparison between results.  In this paper, we present an analysis of
significant sample completeness and contamination issues in high
redshift surveys based on the Lyman break technique, focusing on the
galaxy population at $z\approx5$ to illustrate the complexity of the
situation.  We discuss the points of comparison between different
samples, and the applicability of survey-specific results to the
population at $z>5$ in general, pulling together the modelling and
observational results to construct a coherent picture of the high
redshift population under examination.  An understanding of these issues
is essential to allow fair comparison between different datasets,
between data and theory and when planning future surveys.

In section \ref{sec:completeness} we consider the selection effects
that can influence a high redshift survey and in section
\ref{sec:contamination} we describe the contaminant populations that
also show dropout colours. In section \ref{sec:interpretation} we
discuss the consequences of these effects on the interpretation of
published results from high redshift surveys, and the implications for
the design of future surveys.

\begin{figure*}
\includegraphics[width=0.95\columnwidth]{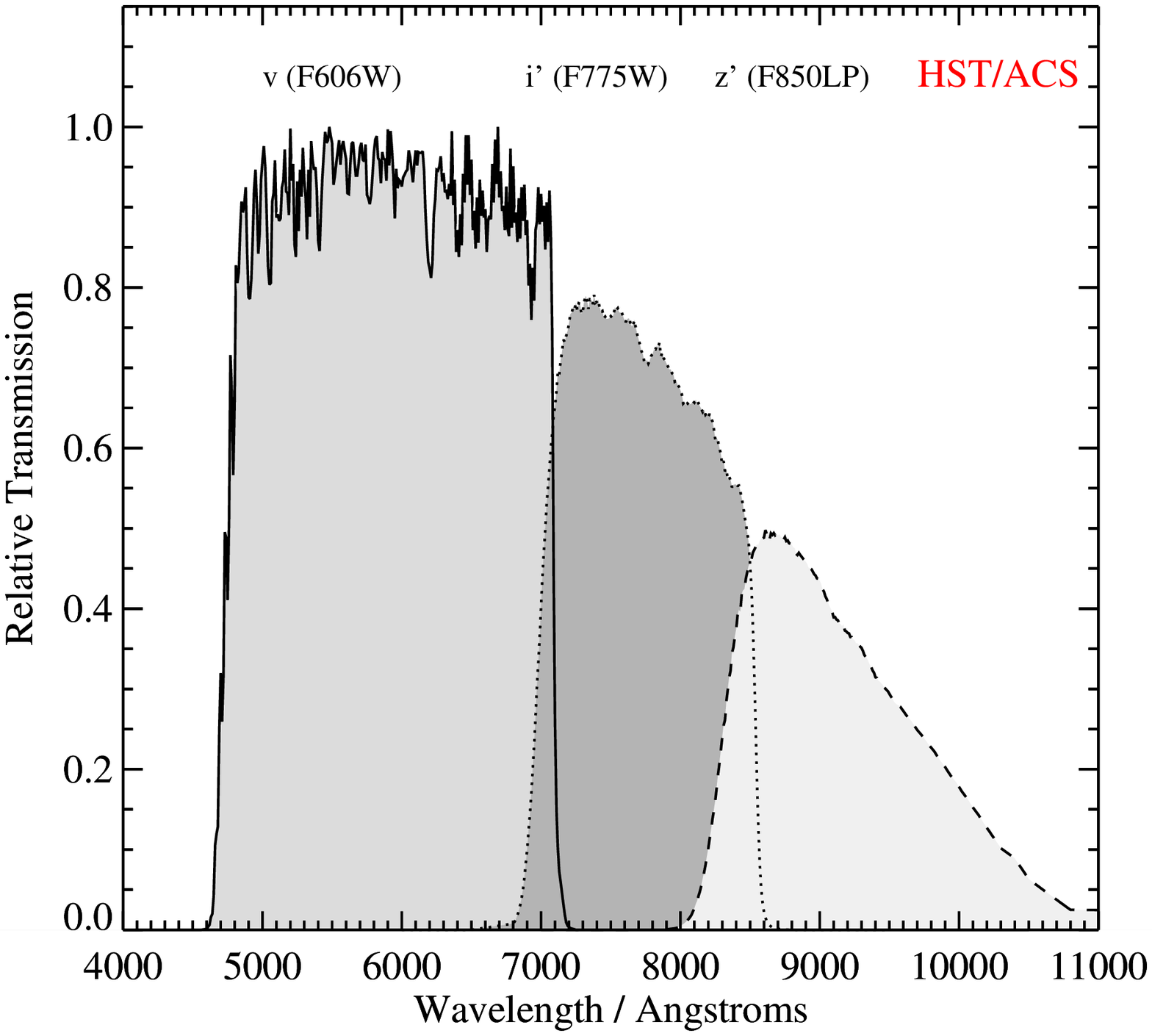}
\includegraphics[width=0.95\columnwidth]{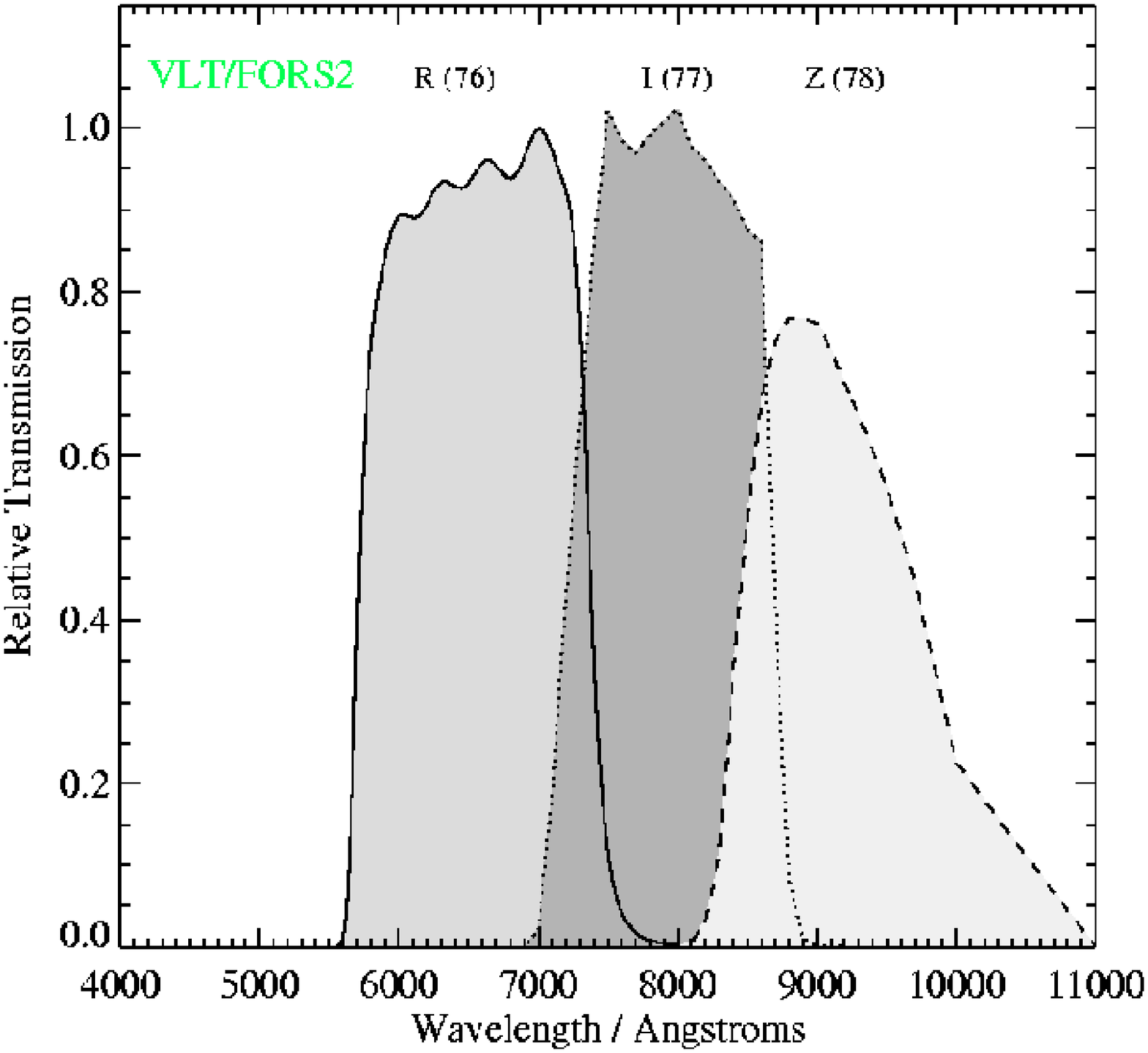}
\includegraphics[width=0.95\columnwidth]{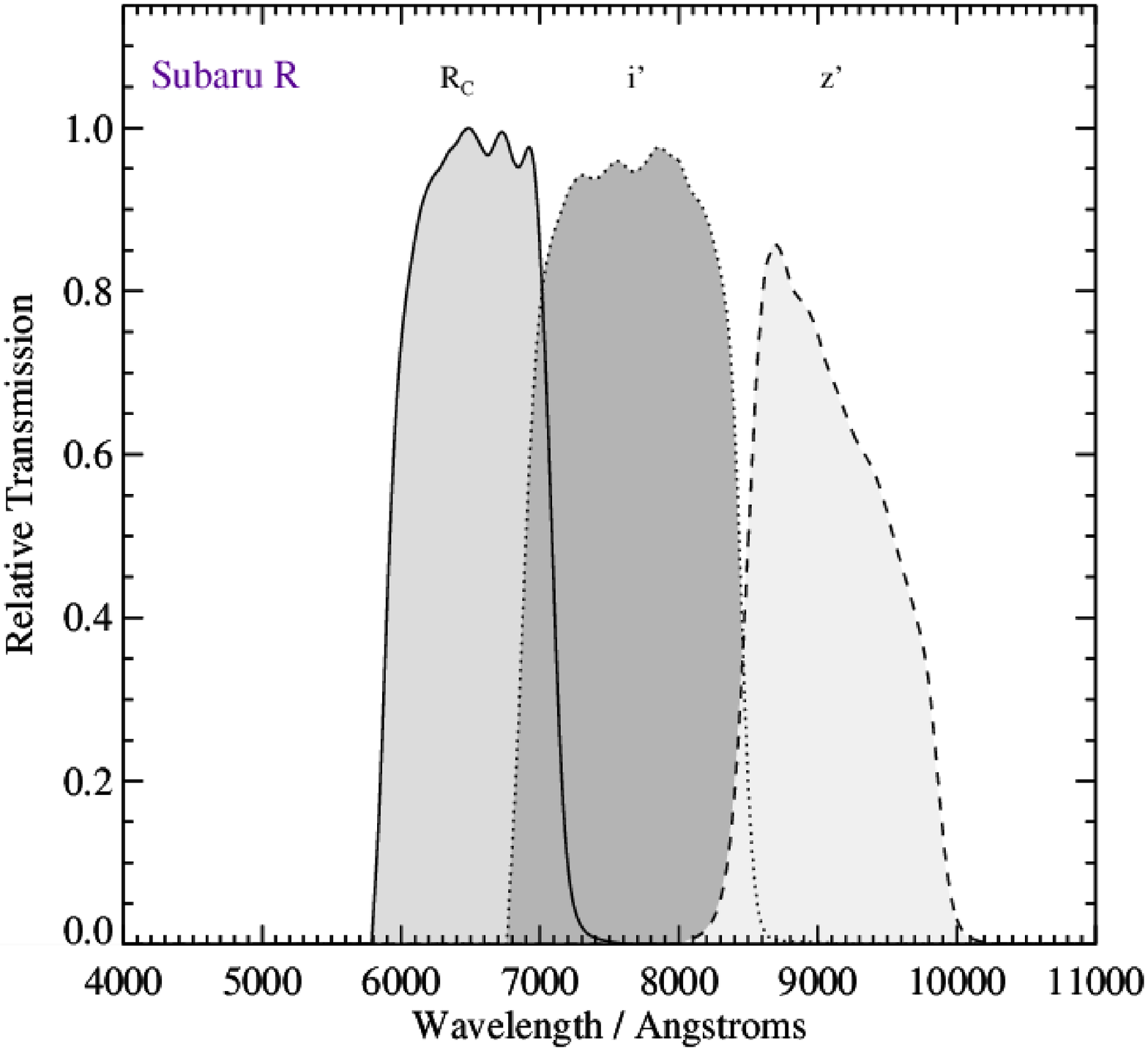}
\includegraphics[width=0.95\columnwidth]{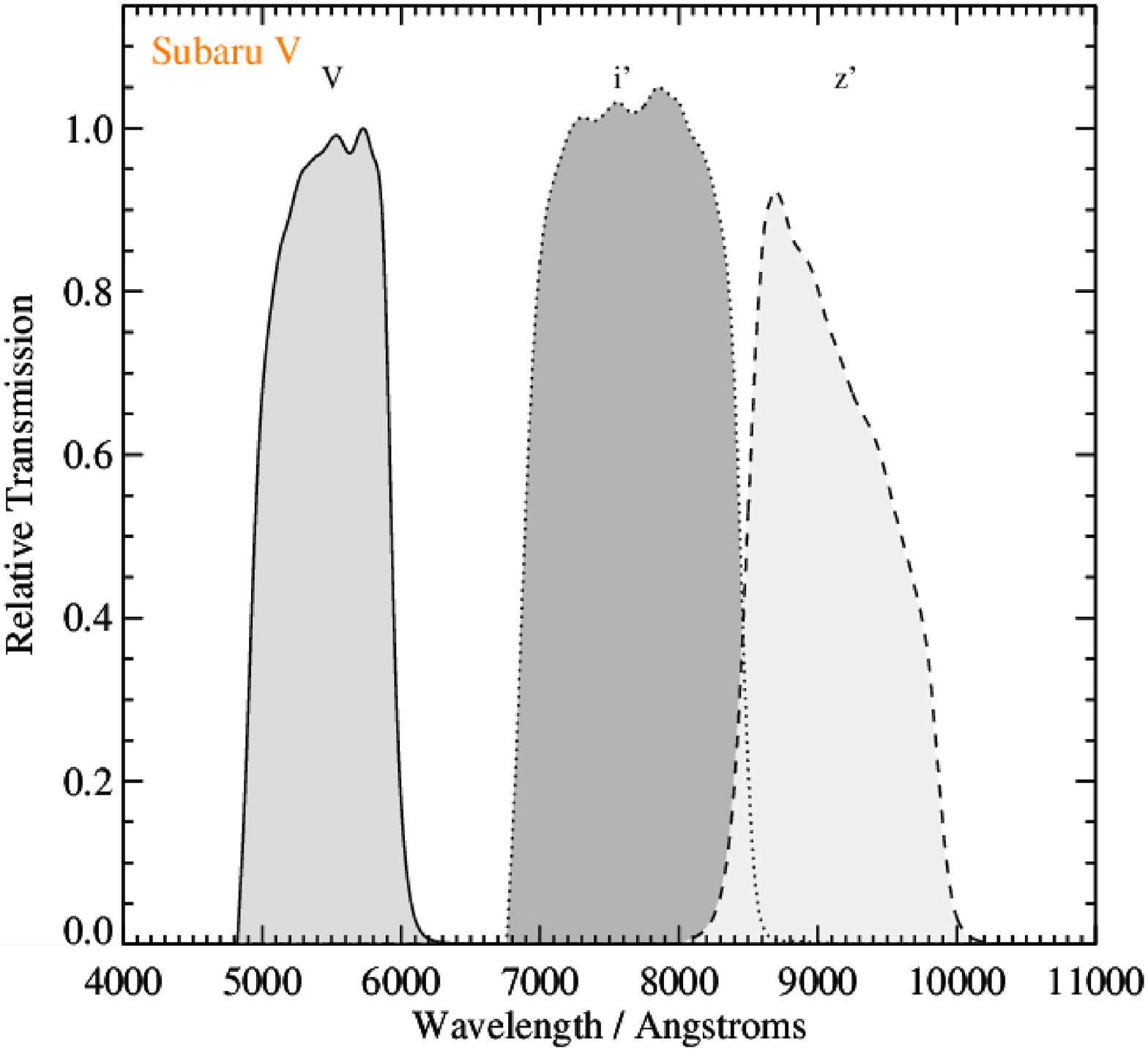}
\includegraphics[width=0.95\columnwidth]{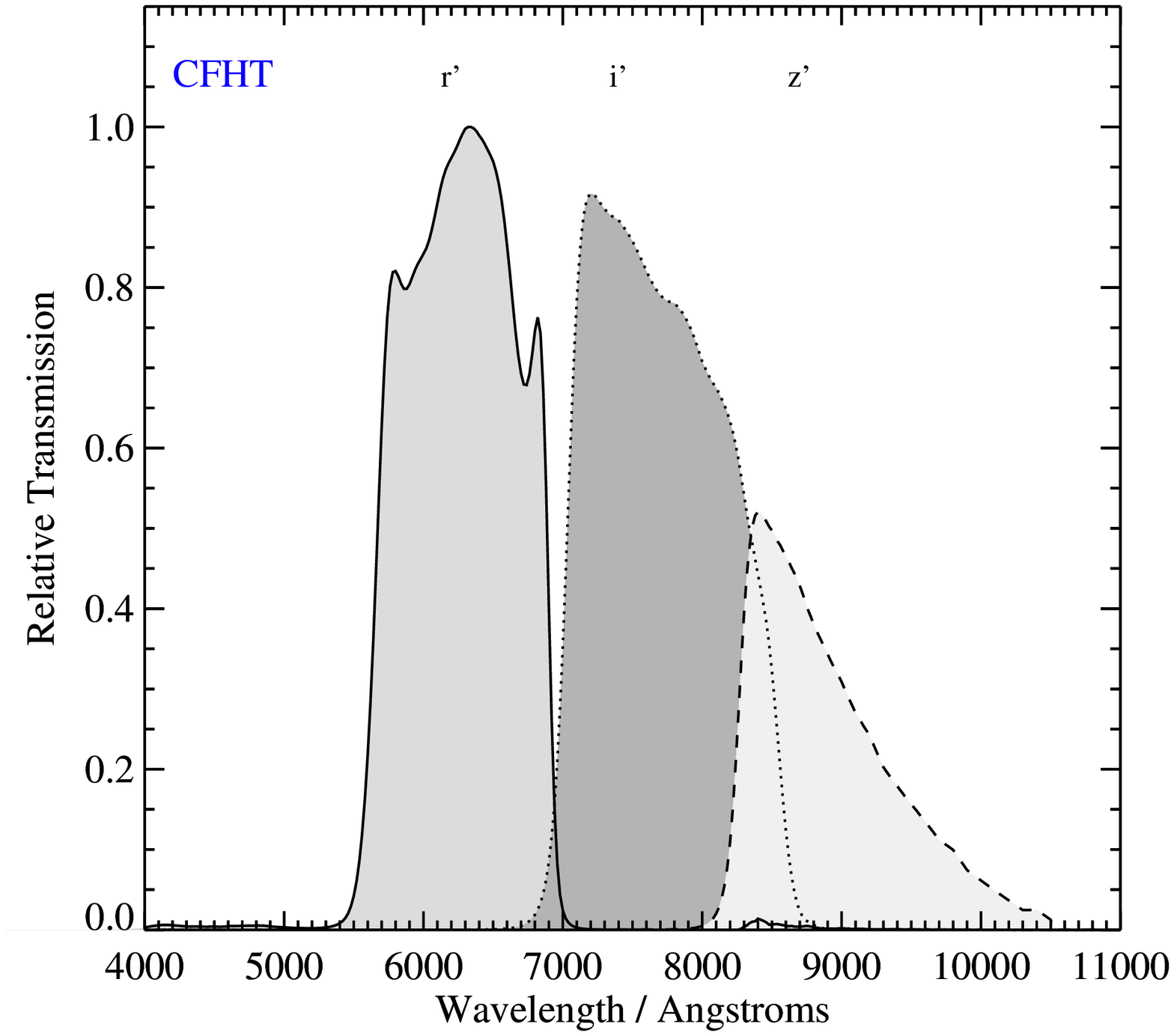}
\caption{The filter sets considered in this paper. In each case the published filter transmission function is convolved with
the appropriate instrumental and CCD response to obtain the final curves.
  \label{fig:filters}}
\end{figure*}

\begin{figure*}
\includegraphics[width=0.95\columnwidth]{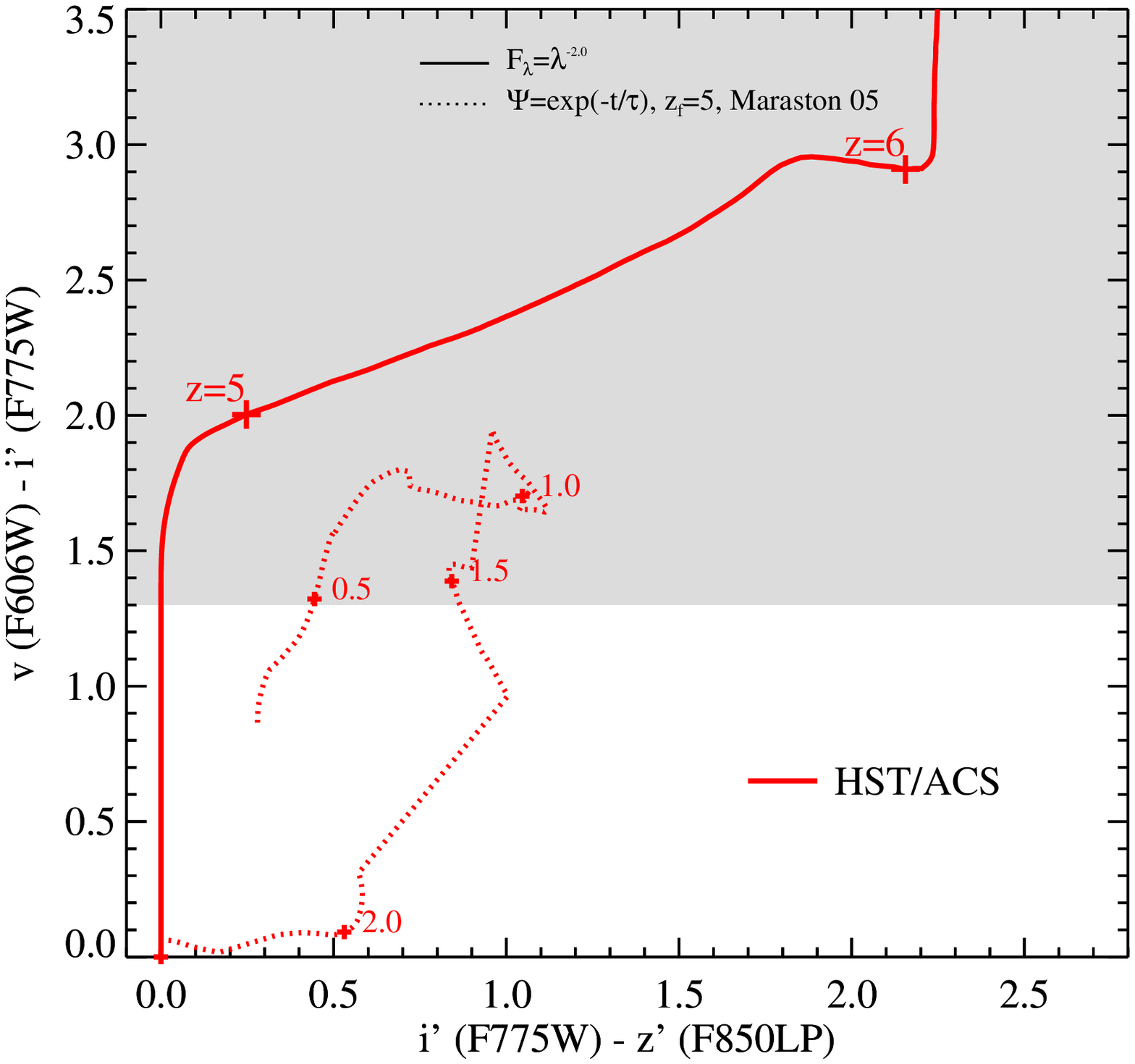}
\includegraphics[width=0.95\columnwidth]{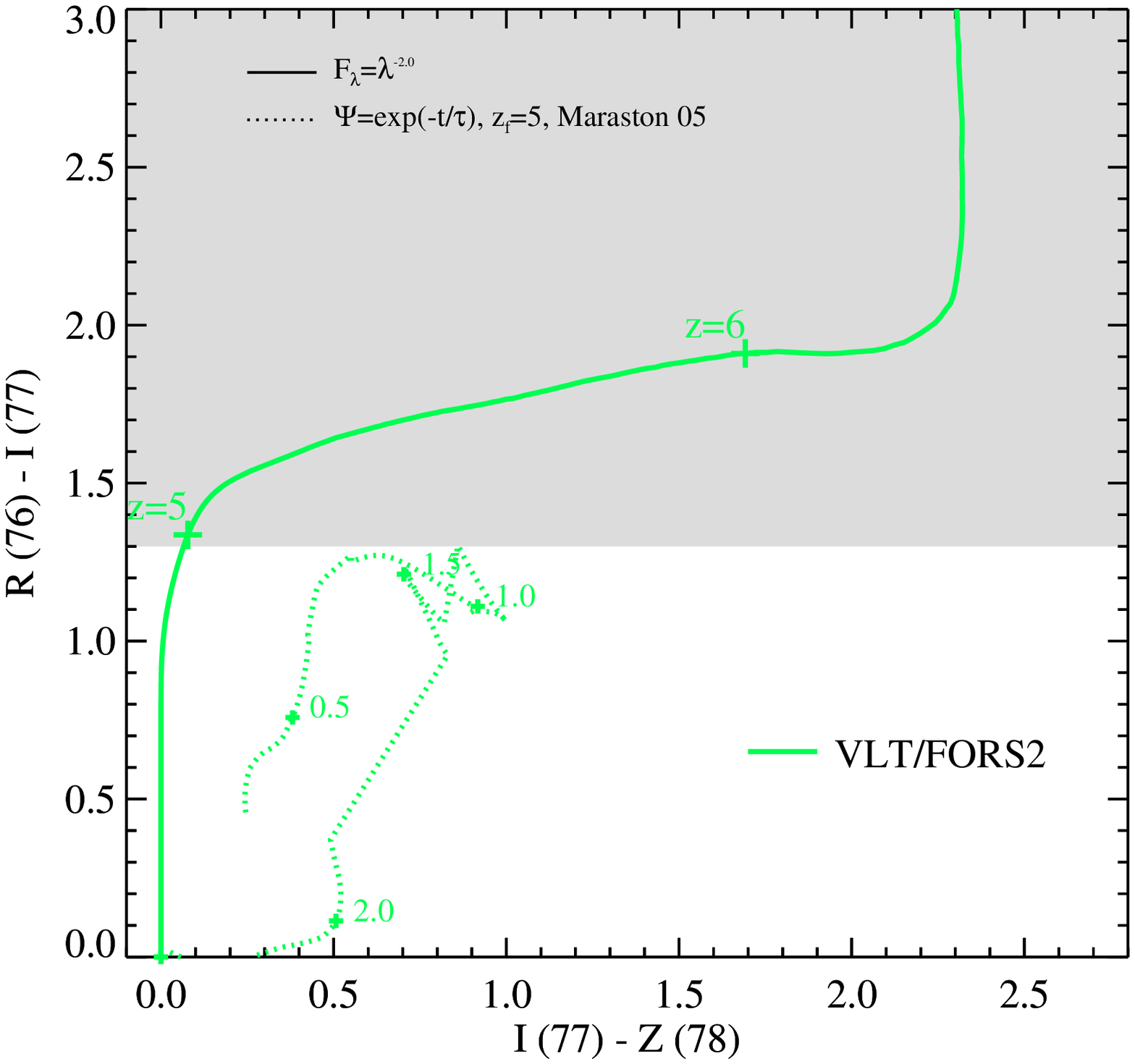}
\includegraphics[width=0.95\columnwidth]{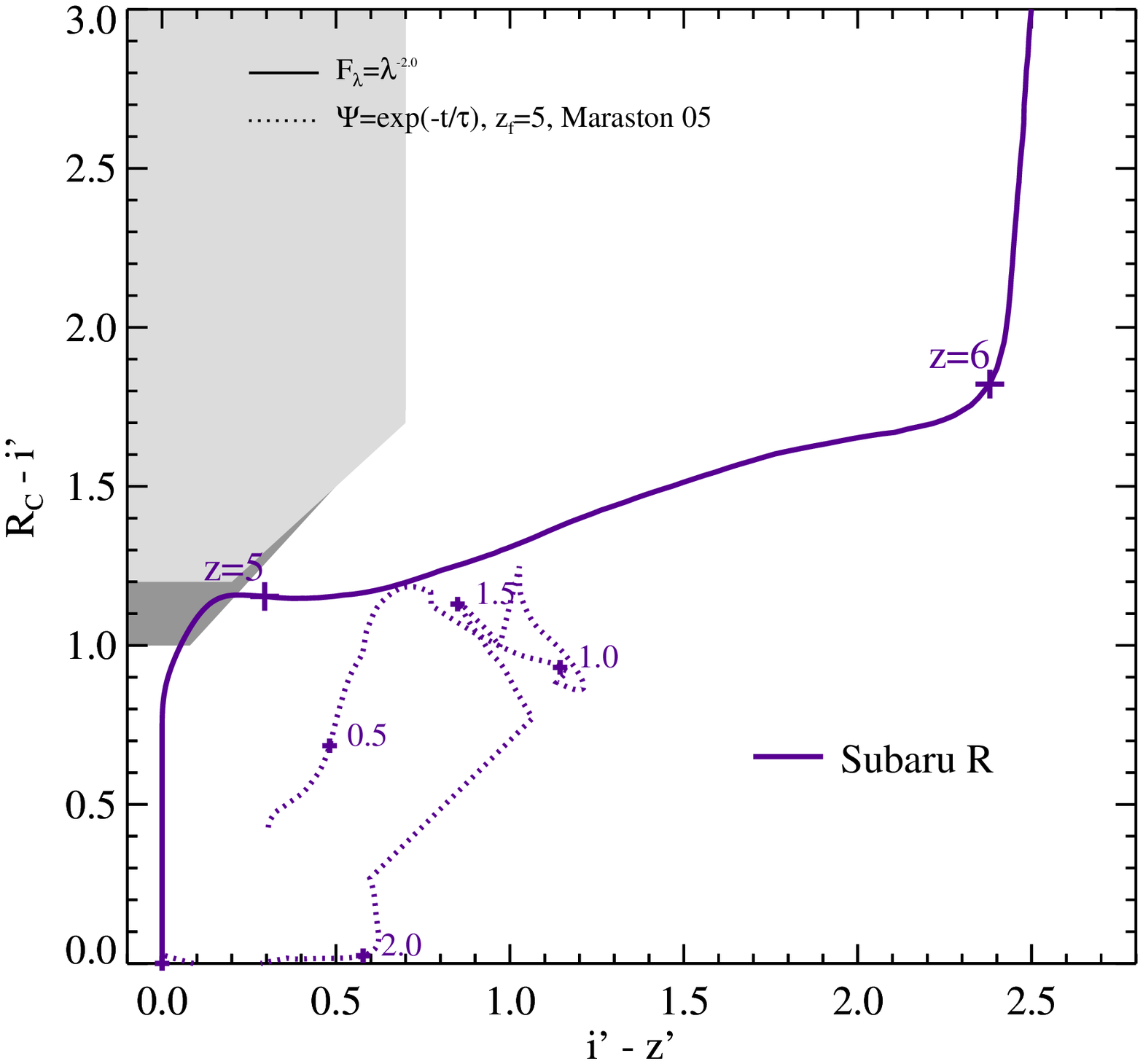}
\includegraphics[width=0.95\columnwidth,viewport=0 0 566 510,clip]{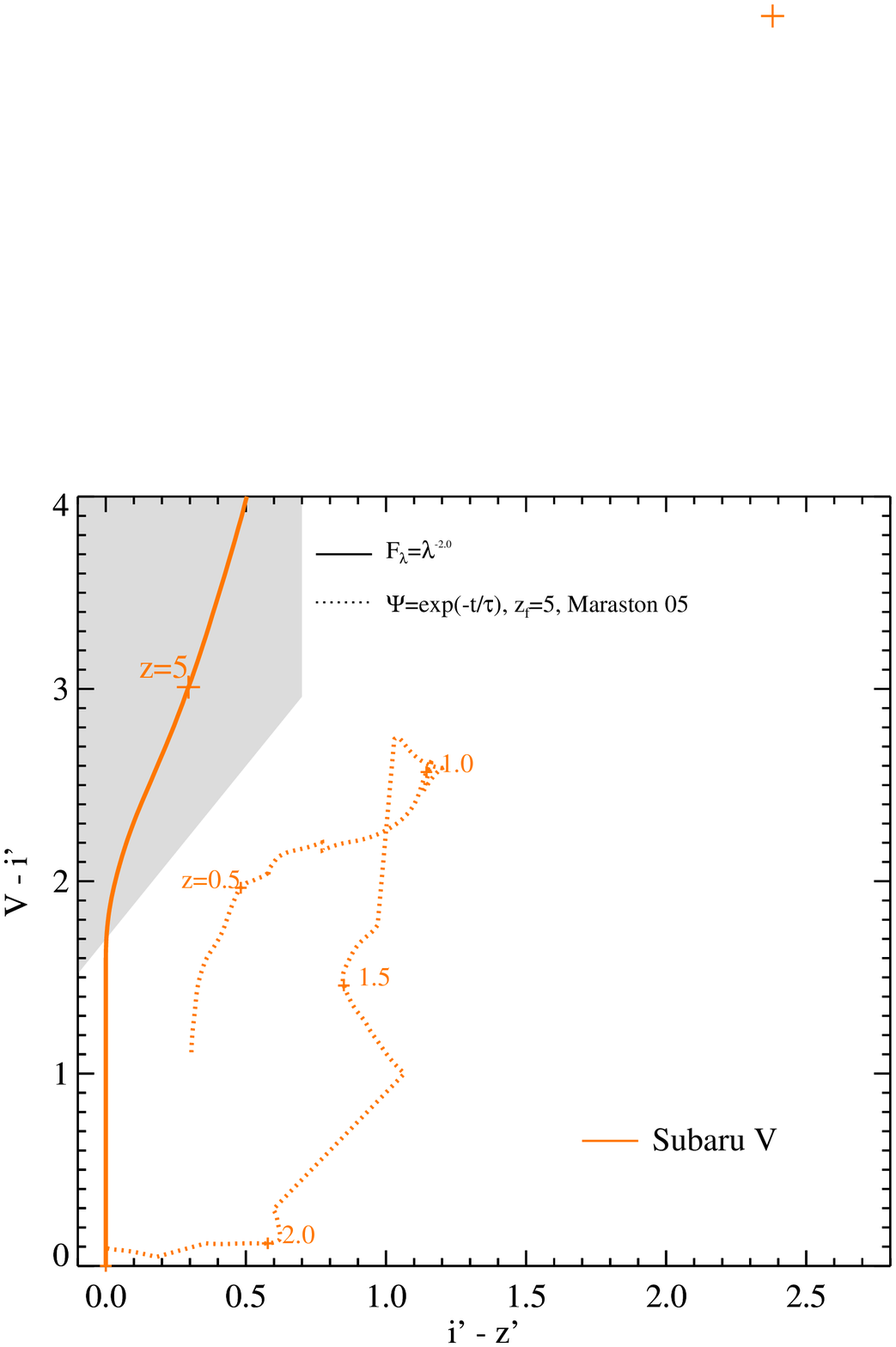}
\includegraphics[width=0.95\columnwidth]{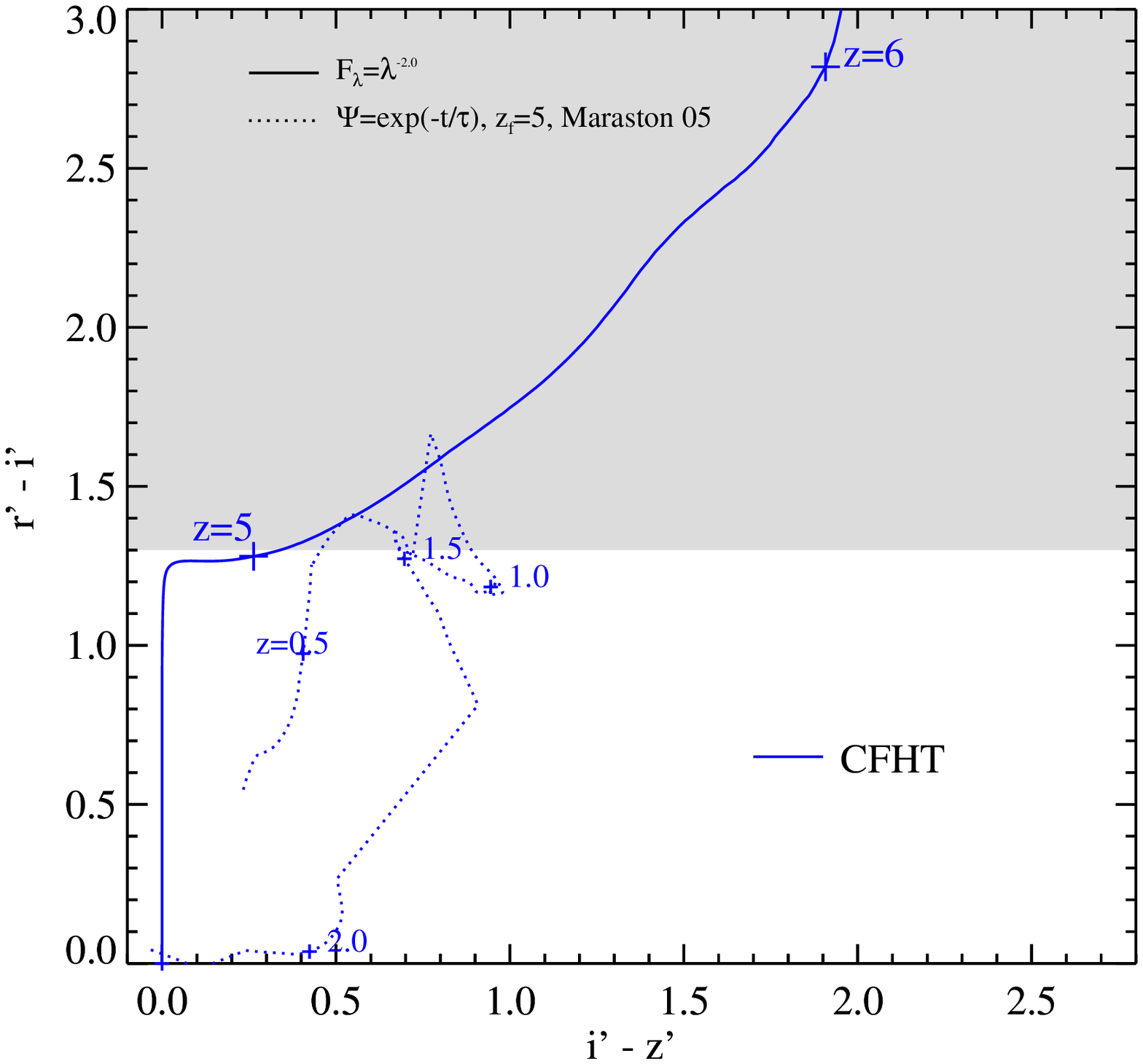}
\caption{The colour selection criteria considered in this paper.
  The colours expected of a flat continuum source (i.e
  $f_\lambda\propto\lambda^{-2}$) at high redshift and a mature galaxy
  at intermediate redshift are shown for reference in solid and dotted
  lines respectively. The high redshift sources are flat in $f_\nu$.
  The intermediate redshift interlopers are constructed from the
  population synthesis models of \citet{2005MNRAS.362..799M}, assuming
  a formation redshift of $z=5$ and a star formation timescale
  $\tau=$0.5\,Gyr. Variation in star formation histories causes
  scatter in the low redshift galaxy locus.  The colours shown in
  these plots are used consistently throughout the paper to represent
  each filter set. The dark shaded region on the Subaru R plot shows
  the more liberal selection criteria of \citet{2006ApJ...653..988Y}
  relative to \citet{2004ApJ...611..660O} (pale shading).
  \label{fig:cols_filters}}
\end{figure*}

Throughout the paper, we consider filter sets and selection functions
that have been applied to existing $z\approx5$ Lyman break galaxy
surveys.  The instruments and filter sets considered, together with
examples of relevant surveys are shown in table \ref{tab:samples} and
illustrated in figures \ref{fig:filters} (filter profiles) and
\ref{fig:cols_filters} (colour selections).  While these selection
functions do not form constitute an exhaustive list, they are
indicative of the variation from survey to survey\footnote{Note, we
  don't discuss the UKIDSS selection of \citet{2006MNRAS.372..357M} in
  detail since the $R-Z>3$ colour cut excludes $z<5.5$ galaxies and
  renders this more properly an $i'$-drop selection. Hence the UKIDSS
  survey expects sources at much lower surface densities making it not
  comparable to other $z\approx5$ surveys.}.  Where appropriate, we
adopt the following cosmology: a flat Universe with
$\Omega_{\Lambda}=0.7$, $\Omega_{M}=0.3$ and $H_{0}=70 h_{70} {\rm
  km\,s}^{-1}\,{\rm Mpc}^{-1}$. All magnitudes (optical and infrared)
are quoted in the AB system \citep{1983ApJ...266..713O}.

\begin{table*}
\begin{tabular}{lp{0.2\textwidth}p{0.25\textwidth}p{0.30\textwidth}}
Facility &  Filters Used & Selection Criteria & Examples \\
\hline\hline
HST/ACS            & {F435W($b$)}, {F606W($v$)}, {F775W($i'$)}, {F850LP($z'$)}
   & $v-i'>1.3$ & \citet{2004MNRAS.347L...7B,2007MNRAS.tmp..294V}\\
VLT/FORS2          & $B$(ESO 74), $R$(ESO 76), {$I$(ESO 77)}, {$Z$(ESO 78)}
   &  $R-I>1.3$ & Douglas et al (in prep), \citet{2003ApJ...593..630L}\\
Subaru/Suprime-Cam & $B$, $V$, $R_c$, $i'$, $z'$ 
   &  {$V-i'>1.2$}, {$i'-z'<0.7$} \& \newline {$V-i'>1.8(i'-z')+1.7$} & `Viz': \citet{2004ApJ...611..660O}, \citet{2006ApJ...653..988Y} \\
  & &  {$R_c-i'>1.2$}, {$i'-z'<0.7$} \& \newline {$R_c-i'>1.0(i'-z')+1.0$} & `Riz': \citet{2004ApJ...611..660O}, \citet{2006ApJ...653..988Y} \\
CFHT/Megacam       & $g'$, $r'$, $i'$, $z'$
   & $r'-i'>1.3$ & None as yet\\
\end{tabular}
\caption{The filter sets and selection criteria examined in this study.
 The last column gives examples of high redshift studies using the selection
 criteria in question. Filter profiles are convolved with the CCD sensitivity 
at the given facility. \citet{2006AA...454..423V} used VLT/FORS2 for spectroscopy.
  The VLT/FORS2 ERGS survey will be fully described in a forthcoming paper by
 Douglas et al and has already yielded one $z=5.4$ AGN \citep{2007MNRAS.376.1393D}
 and dozens of spectroscopically confirmed sources at $z=5-6$. \citet{2003ApJ...593..630L}
 used a more restrictive $R-I>1.5$ colour cut but the same VLT/FORS2 dataset.
  \citet{2006ApJ...653..988Y} slightly modified their selection criteria to allow bluer
  objects than \citet{2004ApJ...611..660O} as shown in figure \ref{fig:cols_filters}.
\label{tab:samples}}
\end{table*}

\section{Completeness Issues in Dropout Samples}
\label{sec:completeness}

\subsection{Redshift Completeness and Number Counts}
\label{sec:filter-effect}

The majority of $z>5$ surveys published in the literature fall into
two categories.  Sources are selected either for the presence of a
strong emission line, detected as excess emission in narrowband
imaging\citep[not discussed further here,
see][]{2006A&A...460..681H}, or by the break in their continuum
emission. In the latter case, the selection is based on an extreme
colour in a single pair of filters, often refined by constraints in
one or more additional bands.  In the case of galaxies at $z\approx5$,
the selection criteria applied in most surveys is either based on
$R-I$ colour or on $V-I$, depending on available imaging, with a
constraint on $I-Z$ often applied \textit{a posteriori} in order to
reduce sample contamination (see section \ref{sec:contamination}).

The requirement for an extreme colour sets a firm lower limit to the
redshift of a survey, while the width of the selection filters defines
the redshift range.  In theory, the simplicity of this approach should
allow samples selected using the same colour cuts on different
instruments to be compared directly.

However such comparisons overlook one significant factor: not all
filter sets are equivalent. There has been a proliferation of filter
sets in the optical, each designed to optimise source signal to noise,
but approaching the problem in different ways.  Such filters are
assigned common names of $V$, $R$, $I$ and $Z$ (or variants on these)
on the basis of their effective wavelength, but often differ
significantly in terms of width, filter response\footnote{combining
  filter transmission profile with available information on CCD
  and instrumental response.} and overlap with
neighbouring filters. This has little effect on the measured
magnitudes in each band of sources with smooth profiles, but can
significantly impact the measured colours of spectra dominated by
sharp features such as continuum breaks.

\begin{figure}
\includegraphics[width=0.95\columnwidth]{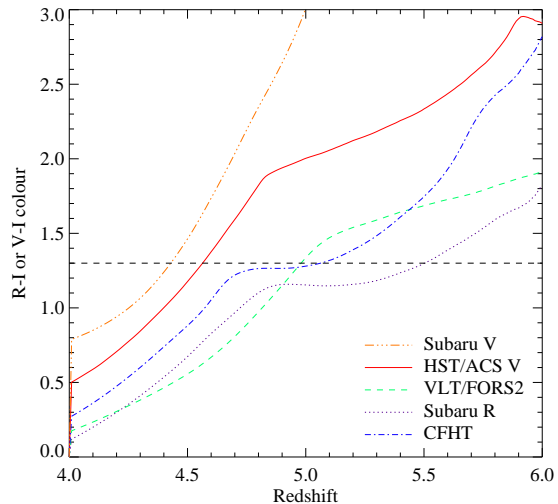}
\caption{The effect of filter profile on redshift selection
  function.  A uniform colour cut of ``R-I or V-I$>$1.3'' will select
  galaxies with minimum redshifts between $z=4.4$ and $z=5.5$
  depending on the filter set in question. Also, due to filter
  overlaps, the colour can flatten out as a function of redshift in
   some redshift ranges leading
  to an increased effect from photometric scatter and intrinsic variation of
  spectral slope around the selection colour.
  \label{fig:cols_rdrops}}
\end{figure}

As figure \ref{fig:cols_rdrops} illustrates, a single high redshift
galaxy \citep[in this case modelled as a source flat in $f_\nu$,
appropriate for young starbursts, and modulated by the intergalactic
hydrogen opacity as a function of redshift given
by][]{1995ApJ...441...18M}, can have dramatically different colours in
different filter sets. A single colour cut simply cannot be applied
uniformly to different filter sets since the resulting redshift ranges
can differ by $\Delta z=1$, particularly if $V$-drops are compared
with sources selected as $R$-drops.

This simple fact is widely understood in the observational community
and most surveys account for this effect by varying their selection
criteria accordingly as shown in table \ref{tab:samples} to tune the
minimum redshift satisfying the colour selection. Nonetheless, the
redshift range probed still varies from sample to sample due to the
width of, and overlap between, the filters available to them.

The most dramatic effect of this redshift variation from survey to
survey arises not from the redshift evolution of the population
itself, but rather from the simple change in luminosity distance over
the redshift range in question. For a magnitude-limited sample, a
progressively brighter limiting luminosity is reached in each
successive redshift slice.  Given the relatively steep faint-end
luminosity function observed for Lyman break galaxies at $z\approx3$,
a small reduction in survey depth at any given redshift can have
dramatic effects on the number of galaxies predicted in that bin.

\begin{figure}
\includegraphics[width=0.95\columnwidth]{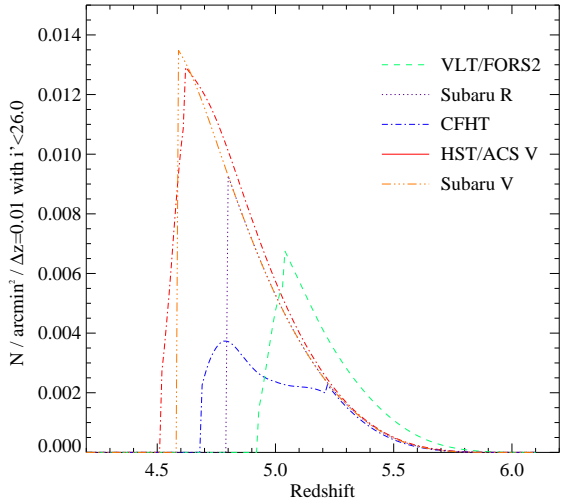}
\caption{The redshift distributions of $z>4.5$ sources,
  selected using different filter sets and the selection criteria of
  appropriate surveys.  Both the surface density of sources and their
  redshift ranges depend sensitively on the selection criteria and
  filter shapes. Counts are plotted assuming the $z=3$ Lyman Break
  galaxy luminosity function of \citet{1999ApJ...519....1S}, and as
  such overpredict the observed number counts at $z\sim5$, and should
  be viewed as indicative rather than predictive. Luminosity function
  effects are discussed in the text. A typical scatter of $\pm$0.15
  magnitudes is assumed, due to scatter in both the intrinsic colours
  and the photometry at these faint limits. The input spectrum is a
  source flat in $f_\nu$. As discussed in section
  \ref{sec:line-emission} and figure \ref{fig:riz_slope}, changing the 
  spectral slope (for example due to the presence of
  dust) has less effect on the redshift distribution than increasing the
  photometric scatter.
  \label{fig:z_dist_rdrops}}
\end{figure}

The combination of these effects - filter response, subsequent
redshift selections and the luminosity bias towards low redshifts - and
their effect on predicted number counts is illustrated in figure
\ref{fig:z_dist_rdrops}. For a constant luminosity function (i.e.
 a non-evolving population from $z\approx3$) the number of galaxies predicted 
as a function of redshift is shown for each of the selection criteria in 
table \ref{tab:samples}.

While the two $V$-drop samples considered here (the selection
functions generally applied to Subaru and {\em HST}/ACS data) overlap
significantly in redshift with the $R$-drop samples, allowing both
groups to be called `$z\approx5$ Lyman Break galaxy populations', the
surface density of sources observed by such samples is some two times
higher. Similarly, there are clear variations within the $R$-drop
samples.  The `Riz' selection utilised by the Subaru Deep Field (SDF) for
example, will identify sources at lower redshifts than that applied by
the ERGS survey at the VLT, but will miss 40\% of Lyman break galaxies
with the same continuum magnitude at $z>5$. 

Also notable is the role played by photometric error in the redshift
distribution.  As figure \ref{fig:cols_rdrops} illustrated, it is
possible for the dropout colour to remain static with redshift in some
filter sets (due either to filter overlaps or to significant offsets
between the $R$ and $I$ filters). If the selection criteria is set
close to such a plateau, then Gaussian scatter in both the intrinsic
colour of the sources and the photometry will promote a fraction of
the more abundant population from lower redshifts into the sample,
while simultaneously scattering a fraction of higher redshift sources
blueward of the limiting colour. The result is a redshift distribution that
bulges below the nominal selection cutoff, as seen in figure
\ref{fig:z_dist_rdrops} for both the VLT/FORS2 and CFHT/MegaCam
selection criteria. It is interesting to note that this effect will be
present for any selection criteria based largely on a single colour, affecting
Lyman-break galaxies across a range of redshifts, as well as the Distant Red
Galaxy population \citep[DRGs, ][]{2003ApJ...587L..79F} 

Figure \ref{fig:z_dist_rdrops} is plotted for the luminosity function
determined by \citet{1999ApJ...519....1S} for Lyman break galaxies at
$z=3$. This luminosity function is known to overpredict the number of
$z>5$ sources of bright magnitudes \citep{2003ApJ...593..630L}, and
hence the predicted number counts in figure \ref{fig:z_dist_rdrops}
should be considered indicative rather than precise. However, while
the normalisation of figure \ref{fig:z_dist_rdrops} may change, the
basic differences in the redshift distributions of sources will remain
unless the shapes of those distributions also change.

Altering the luminosity function
parameters has no effect on the colours of a high redshift source
(which are fixed by rest-frame UV slope and Lyman-$\alpha$ forest
transmission), but does effect the number of sources in a given volume
which will satisfy this criterion. Decreasing the typical luminosity
L* to half its value at $z=3$ reduces the peak number density of
sources by a factor of six, but affects the shape less severely,
broadening the redshift distribution at FWHM by approximately 10\%.
Similarly, increasing the faint end slope $\alpha$ from -1.5 to -1.9
has the effect of decreasing the peak number density by 15\% while
leaving the shape and FWHM of the distribution unchanged.

Hence the primary effect is on the normalisation rather than the shape
of the redshift distribution and the filter-to-filter comparison
discussed here are largely independent of luminosity function.

The effect of changing the limiting luminosity of a survey is rather
more pronounced. Such a change does not alter the minimum redshift
identified by the survey but does increase the depth relative to L* in
each successive redshift bin.  Since the luminosity function is steep
at the faint end, this has the effect of broadening the redshift
distribution in a given filter set and extending the tail of the
function to higher redshifts.


\subsection{Effect of Line Emission}
\label{sec:line-emission}

Lyman break galaxy surveys focus on the analysis of galaxies detected
through their rest-frame ultraviolet continua. However, it is known
from Lyman-$\alpha$ emitter surveys
\citep[e.g.][]{2006ApJ...638..596A,2004AJ....127..563H,2004ApJ...617L...5M}
that there exists a significant population of sources with powerful
emission lines at $\lambda_\mathrm{rest}=1216$\AA, a fraction of which
overlaps with the Lyman break galaxy population. At $z\approx3$, a
quarter of all spectroscopically confirmed Lyman break galaxies show
Lyman-$\alpha$ emission with a rest-frame equivalent width of
$W_0>20$\AA, while a second quartile has $0<W_0<20$\AA\
\citep{2003ApJ...588...65S}.

\begin{figure}
\includegraphics[width=0.95\columnwidth]{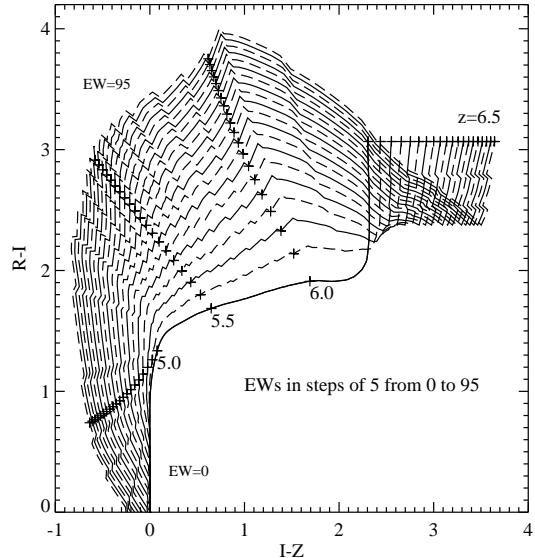}
\caption{The effect of moderate-strength Lyman-$\alpha$ line emission
  on the R-I, I-Z colours of high redshift galaxies, illustrated in
  the ESO filter set used at the VLT. Rest-frame equivalent widths of
  zero to 95\AA\ are shown from lower-right to upper-left in
  increments of 5\AA, imposed upon a constant continuum flux. At $z<5$
  the emission line is in the $R$-band, while at $5<z<6$ the emission
  line is in the $I$-band and hence affects both colours. Above $z=6$,
  the emission line enters the $Z$-band, before leaving the filter set
  entirely. Colours are for a source with constant $F_\nu$. If strong
  line emitters have blue spectral slopes, the effect would be more 
  pronounced still (see figure \ref{fig:riz_slope}).
  \label{fig:riz_ew}}
\end{figure}

Figure \ref{fig:riz_ew} illustrates the effect of moderate line
emission on the colours of a flat-spectrum continuum-source, for one
example filter set - in this case, the VLT/FORS2 filters. As in most
surveys, a flat-spectrum galaxy will satisfy the VLT/FORS2 selection
criterion at redshifts where the Lyman-$\alpha$ feature is in the
$R$-band.  While this remains true, increasing line strength has the
effect of reducing the galaxy colour and dropping it out of the
sample at the low redshift end of the selection function.
At higher redshift, the presence of a line-emitting population has the
effect of scattering galaxies over a much broader region than the
simple galaxy locus, potentially reducing the ability of a survey to
reliably separate galaxy loci from contaminants (see section
\ref{sec:contamination}). 

A second effect of line emission is to increase the observed magnitude
of a galaxy when compared to an equivalent continuum source without
line flux. At faint magnitudes, a moderate strength emission line can
contribute a large fraction of the observed broadband flux,
particularly at high redshifts (since only a small fraction of the
filter is free of Lyman-$\alpha$ forest line dampening). Given the
steep faint-end slope expected for any reasonable luminosity function,
the population of galaxies just beyond the (continuum) magnitude limit
of a survey is larger than the faintest population above the limit.
Hence a small fraction of those galaxies, entering the sample due to
the presence of strong emission lines, can skew both redshift and
equivalent width distributions.

\citet{2007MNRAS.376..727S} recently found evidence for a tail of high
equivalent width sources in a faint Lyman break galaxy sample at
$z\sim6$ from the Hubble Ultra Deep Field \citep{2006AJ....132.1729B},
suggesting that the effect of strong line emission on the tail of the
galaxy distribution could be measurable in a larger sample at
equivalent faint limits.

Each of these effects will be dependent on the position of
Lyman-$\alpha$ emission relative to the filter edges, and also on the
detailed profile of those filters.  A filter with a square-edged
transmission profile will have clear advantages in terms of reducing
the tail of galaxies that are detected in spectral regions with very
little filter throughput.  However the decline in throughput of
optical CCDs through the $Z$-band leads to an inevitable blurring of
the edges, and square-transmission filters will not prevent effects arising from
incomplete blanketing of one or more bands by Lyman-$\alpha$ forest
absorption.


\subsection{Effect of Spectral Slope and Burst Age}
\label{sec:stellar_pop}

\begin{figure}
\includegraphics[width=0.95\columnwidth]{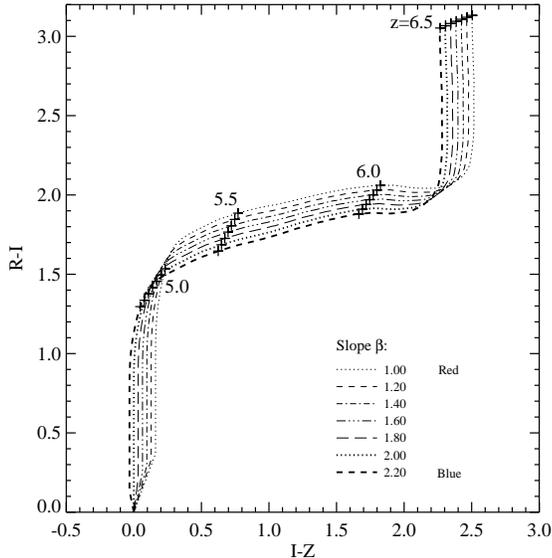}
\caption{The effect of varying the rest-frame ultraviolet slope, defined
  as $f_\nu\propto\lambda^{-\beta}$, on the R-I, I-Z colours of high
  redshift galaxies, illustrated in the ESO filter set used at the
  VLT. A source with constant $F_\nu$ has a spectral slope
  $\beta=2.0$. Sources with very blue spectral slopes will be lost
  close to the selection boundaries, although this effect is small
  compared with that due to emission lines (figure \ref{fig:riz_ew}).
  \label{fig:riz_slope}}
\end{figure}

By contrast, intrinsic spectral slope has a relatively small effect on
the colours of high redshift galaxies, which are dominated instead by
the effects of interstellar absorption.  Nonetheless, a blue
rest-frame ultraviolet spectral slope (appropriate for young sources)
might reasonably be associated with strong line emission, and so will
strengthen the effect seen above.  The effect of varying the
ultraviolet spectral slope (in the absence of line emission) is shown
in figure \ref{fig:riz_slope}. The effects of a bluer average
ultraviolet continuum at $z=5$ \citep[possibly steeper than
$\beta=2.0$, see][]{2005MNRAS.359.1184S} compared with the lower
redshift population \citep[typically
$\beta=1.1-1.5$,][]{1999ApJ...519....1S} are unlikely to have a big
impact on the selection since they are of the same order as
photometric errors at these faint magnitudes, but could produce a
measurable effect for sources close to the selection limits.

The most significant factor driving the rest-UV spectral slope is the
age of the current starburst in a galaxy. The hottest, most-massive
stars provide the largest contribution to the rest-frame ultraviolet,
and these stars also have the shortest lifetimes.  As a result, the
optical colours of galaxies at $z\approx5$ evolve with the age of the
galaxies in question.

\begin{figure}
\includegraphics[width=0.95\columnwidth]{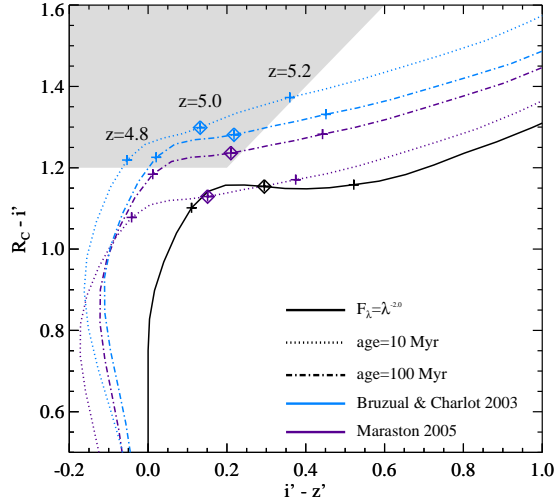}
\caption{The effect of stellar population age on the model colours of 
  high redshift galaxies. Here we show the colours of galaxies forming
  stars continuously over 10 (dotted) and 100\,Myr (dot-dash),
  contrasted with the flat spectrum expected for a young instantaneous
  starburst (solid). The colours of the widely used Bruzual \& Charlot
  (2003) models are shown in pale blue, while those predicted by the
  more recent Maraston (2005) models are in purple.  The Subaru
  $R$-band selection filters and colour criteria are used.
  \label{fig:stell-pop}}
\end{figure}

In figure \ref{fig:stell-pop} we explore the effect of changing the
template SED on the selectability of high redshift galaxies, with
particular reference to the Subaru $R$-band colour selection in table
\ref{tab:samples}. The colours of a spectrum flat in $f_\nu$ is
compared with those predicted by stellar population synthesis models
for galaxies constant star formation over the preceding 10\,Myr and
100\,Myr (assuming solar metallicity and a Salpeter IMF). We consider
the results of two different stellar synthesis codes for identical
output parameters. The \citet{2003MNRAS.344.1000B} codes have been the
most widely used synthesis models in recent years and were utilised by
\citet{2004ApJ...611..660O} to estimate the photometric selection
criteria for high redshift galaxies. They produce redder model colours
at a given redshift than the newer stellar synthesis models of
\citet{2005MNRAS.362..799M}, which incorporate improved treatment of
the thermally-pulsating asymptotic giant branch phase
\citep[see][]{2007astro.ph..2091B}.

Evidence from SED fitting to the stellar populations in $z>5$ galaxies
\citep{2007MNRAS.tmp..294V,2007ApJ...659...84S,2005MNRAS.364..443E}
has determined that the majority of this population comprise either
young starbursts ($<$50\,Myr) or young starbursts with an underlying
older ($>$\,200Myr) population that no longer contributes
significantly to the rest-UV.  In either case, no compelling evidence
has been found for continuous star formation over timescales of
100\,Myr. This youth of stellar populations at high redshift is
supported by other work
\citep[see][]{2003ApJ...593..630L,2005MNRAS.359.1184S,2007A&A...471..433P},
which finds colours consistent with a flat spectrum in bands
uncontaminated by the break in $z>4$ samples.

For the purposes of this paper we adopt a spectrum flat in $f_\nu$
(i.e. $\beta=2.0$) as being a simple template galaxy. As figure
\ref{fig:stell-pop} illustrates, this is appropriate for young
starburst populations \citep[in the][models]{2005MNRAS.362..799M},
although we note that uncertainties in the star formation history can
lead to variation in colour by approximately 0.1\,mag (or less than
the photometric scatter in a typical survey of faint dropouts).

\subsection{UV-faint Galaxies}
\label{sec:intrinsic-colour}

The dropout selection technique, by its very nature, identifies
sources with a bright rest-frame ultraviolet continuum. Ultraviolet
flux is contributed primarily by young, short lived stars - hence the
use of Lyman break selected samples to gauge the star formation
history of the universe
\citep{1996MNRAS.283.1388M,1999ApJ...519....1S,2003ApJ...593..640B,2003MNRAS.342..439S}.
 
However, the ultraviolet continuum flux declines sharply with
increasing time after an instantaneous starburst
\citep{1999ApJS..123....3L}, varying by three orders of magnitude in
the first 100\,Myr.  Hence the Lyman break technique cannot identify
galaxies at high redshift that have passed through a starburst and are
now passively evolving since such sources will no longer have a strong
ultraviolet continuum. At $z\approx5$ a 100\,Myr-old starburst would
have peaked at only $z\approx5.5$ and yet may be undetectable by
dropout selection (unless star formation continues throughout this time).

Similarly, it takes a finite time after a starburst for an
ultraviolet continuum flux generating population to become
established, causing very young starbursts ($<$5\,Myr) to be detectable
primarily in emission from the Lyman-$\alpha$ emission line. 
Hence both very young and very old starbursts may be omitted from a
dropout survey, while galaxies of $10-100$ Myrs in age may be
preferentially selected based on ultraviolet flux. 

Whenever only isolated subsets of the total galaxy population is
observed, it is possible to miss aspects of the bigger picture.
Populations that are small in number - such as the DRG galaxies (a
combination of old and dusty galaxies) at $z\approx2$ - can make
non-negligible contributions to important cosmological parameters such
as the galaxy metal budget \citep{2006MNRAS.367L..16B} and stellar
mass density \citep{2007ApJ...656...42M}.  Hence it is important to
consider constraints on galaxies not selected by high redshift dropout
samples.

An analogue to the passively evolving subset of $z\approx2$ DRGs at
$z\approx5$ would be old sources without significant ongoing star
formation.  In these old galaxies, the most easily detected spectral
feature would be the rest-frame 4000\AA\ break.
\citet{2005ApJ...635..832M} have proposed that infrared data can be
used to detect the Balmer break in $z>5$ galaxies. Although their
initial candidate may well lie at lower redshifts, the technique has
now identified a small number of candidate galaxies at comparatively
bright magnitudes
\citep{2007arXiv0705.0660C,2007astro.ph..3276R,2007MNRAS.376.1054D}.
Unfortunately a degeneracy in observed optical-infrared colour between
$z\approx2$ dusty galaxies and non-starforming $z>4$ galaxies renders
the analysis of such a population based on photometric redshifts alone
difficult. Photometric redshifts in the available broad wavebands can
favour one solution, but not to the exclusion of the other, and
quantitative results cannot be drawn without appropriate caveats and
speculative corrections.  Detailed investigation of high redshift
galaxies without significant rest-UV flux will most likely require a
future generation of telescopes and instruments capable of performing
rest-frame optical spectroscopy on extremely faint sources.

The importance of this population is difficult to quantify.  If the
most massive sources collapsed earliest then old galaxies could
conceivably be amongst the brightest high redshift sources and hence
retain a measurable ultraviolet flux to any given magnitude limit
despite the rapid decline of ultraviolet emission after its initial
peak. Many old galaxies at $z>5$ may also have ongoing star formation
and so examples have been identified as high redshift sources on the
basis of recent starburst activities rather than through their evolved
populations \citep[e.g.][]{2005MNRAS.364..443E}. The population of old
galaxies not-selectable as $V$- or $R$-drops may be constrained through
measurements of star formation at $z>7$, but before that is necessarily
a matter of speculation.

In addition to incompleteness in terms of old, massive galaxies,
Lyman-break samples may also be incomplete for young galaxies of
similar mass to those observed.  \citet{2007MNRAS.tmp..294V} fitted
the optical and ultraviolet spectral energy distributions of
$z\approx5$ $v$-band dropouts, and determined that the typical age of
such galaxies is $\approx30$\,Myr, with approximately one third of the
sample showing evidence for underlying older ($>100$\,Myr) starbursts.
While there is clearly a bias against old galaxies when selecting on
unobscured ultraviolet luminosity, an intermediate age population of
$10-100$\,Myr starbursts should have been easily detected. As
\citeauthor{2007MNRAS.tmp..294V} discuss the presence of such a large
population of short-lived sources distributed throughout the
comparatively long (325\,Myr) time span probed by the sample suggests
that the detected galaxies represent a more passive population an
order of magnitude more numerous that shows stochastic bursts of star formation.

Neither of the above selection effects accounts for the additional
effect of dust extinction, which can suppress the rest-frame
ultraviolet flux. At lower redshifts ($z=1-4$) several species of
dust-obscured galaxies are known including sub-millimetre galaxies and
ultra-luminous infrared galaxies (ULIRGS).  The generally blue
rest-frame UV slope at $z>5$
\citep[e.g.][]{2005MNRAS.359.1184S,2006ApJ...653...53B} suggests that
dust extinction at high redshifts may be lower than those observed at
$z\approx3$. A dust extinction curve derived from observations of
$z>6$ quasars \citep{2004Natur.431..533M} suggests that high-redshift
dust is generated primarily in supernovae and produces up to a
magnitude less extinction in the ultraviolet than the more-processed
dust seen at lower redshift. The same extinction curve provides a good
fit to the spectrum of a gamma ray burst that also lies at $z>6$
\citep{2007astro.ph..3349S}. If these results are typical of galaxies
at high redshifts, then the population lost to dust extinction down to
any given magnitude is likely to be smaller than that at lower
redshifts, although the presence of sub-millimetre galaxy analogues
cannot be ruled out, and mildly dust-obscured dropouts could be a significant
cause of incompleteness at the faint end of any survey.

Sample incompleteness over the sensitive redshift range of Lyman-break
samples is dominated by the unobscured rest-frame ultraviolet flux and
the strength of spectral breaks, and hence are less filter-dependent
than those discussed above. These effects are, however, significantly
harder to quantify since they rely on the existence of a population
that has never actually been observed.

%
%
%
%

\section{Contamination Issues in Dropout Samples}
\label{sec:contamination}

Two distinct populations of astronomical objects
- intermediate redshift elliptical galaxies and cool Galactic stars -
are degenerate in colour with high redshift galaxies.

While many such
contaminants can be distinguished through the use of data in the {\em
  Spitzer}/IRAC bands longwards of 3\,$\mu$m, such imaging is not
always available, and in many cases is impossible to obtain because of
source confusion at long wavelengths. Similarly, a spectroscopically
complete Lyman-break survey can accurately correct for contamination
effects, but this realistically limits such a survey to limits of
$I=26.5$ or brighter on an 8m telescope.

Hence understanding and correcting for these contaminant populations
is vital when interpreting a photometrically selected sample,
particularly at faint magnitudes.

  \subsection{Cool Galactic Stars}
  \label{sec:cool-galactic-stars}
 
\begin{figure*}
\includegraphics[width=0.95\columnwidth]{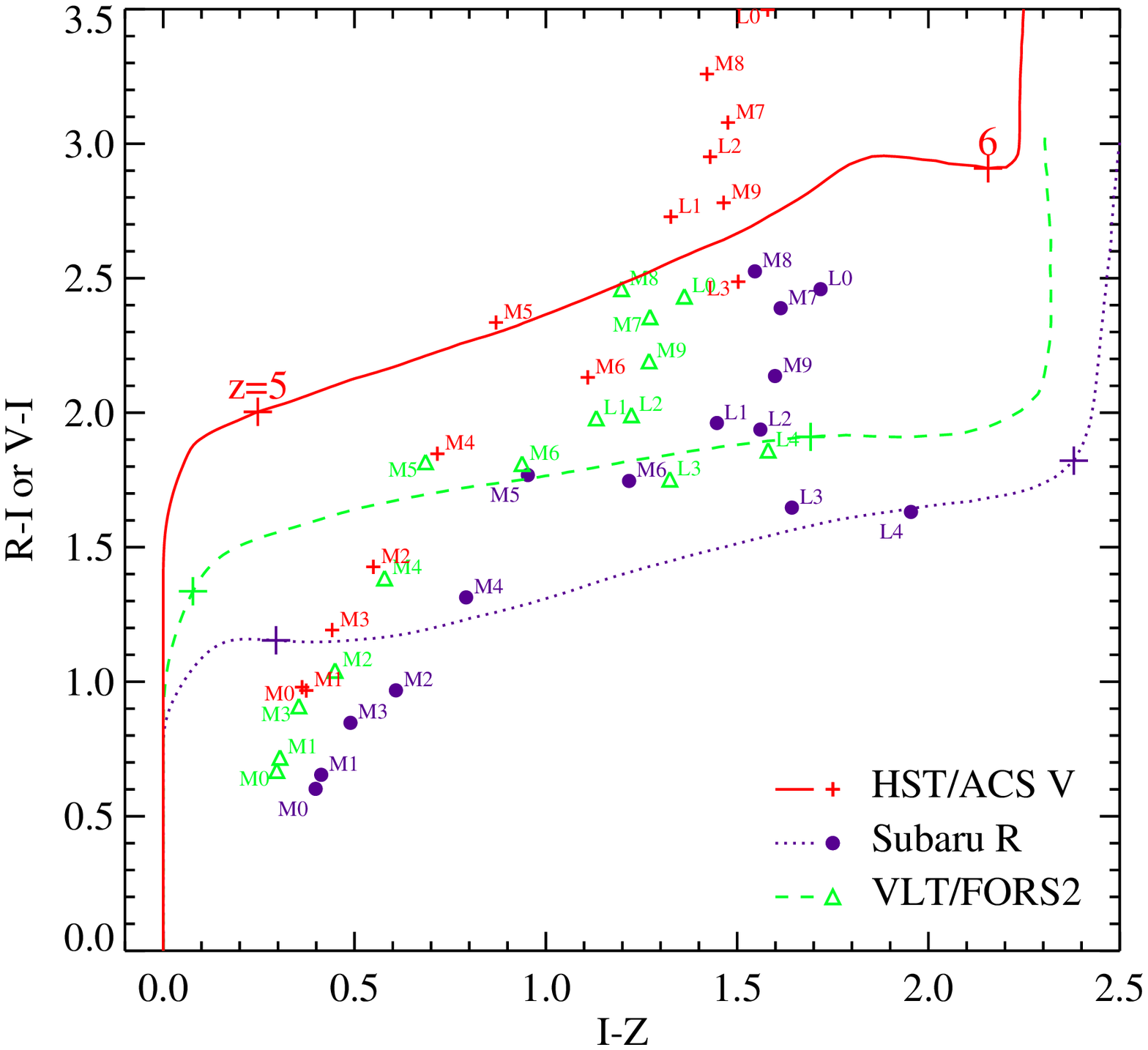}
\includegraphics[width=0.95\columnwidth]{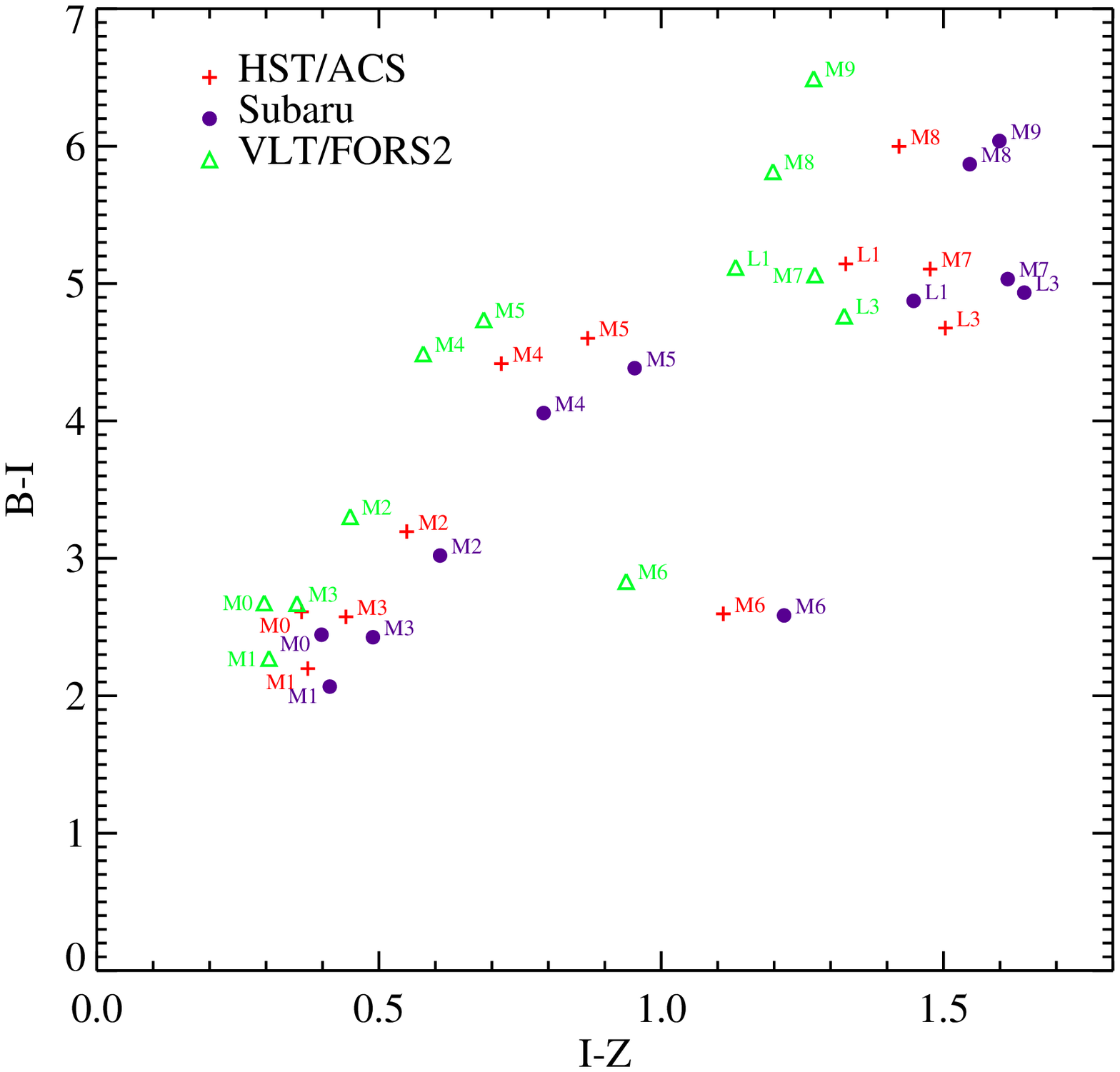}
\caption{The $R-I$ (a, left) and $B-I$ (b, right) versus $I-Z$ colours of observed dwarf star templates
  \citep{2002AJ....123.3409H}, convolved with three commonly-used
  filter sets. In the left hand panel the redshift tracks of a flat
  spectrum galaxy are also shown. The depth of $B$-band imaging
  required to eliminate dropout stars can vary by more than half a
  magnitude depending on filter profile, and exceeds that available in
  most deep surveys.
  \label{fig:biz_stars}}
\end{figure*}

  Cool galactic stars of classes M4 and later are routinely selected
  in dropout selections. M class stars satisfy $V$ or $R$-drop
  colours, while L and T stars have $I$-drop colours (figure
  \ref{fig:biz_stars}).
  
  In fields with {\em HST}/ACS imaging, stars are routinely excluded
  on the basis of their unresolved full-width
  half-maxima\footnote{While this risks omitting compact galaxies, all
    confirmed $z>5$ galaxies thus far observed from space are
    resolved.}. However this approach cannot be used in wide-field
  imaging, since all known (unlensed) $z>5$ galaxies are unresolved
  from the ground. Instead the usual approach adopted \citep[e.g. the Subaru
  Deep Field studies, ][]{2004ApJ...611..660O} attempts to
  exclude stars on the basis of detected flux shortwards of the
  nominal Lyman-$\alpha$ break or through placing constraints on a
  second colour such as $I-Z$.
  
\begin{figure}
\includegraphics[height=0.90\columnwidth,viewport=0 0 566 566,clip]{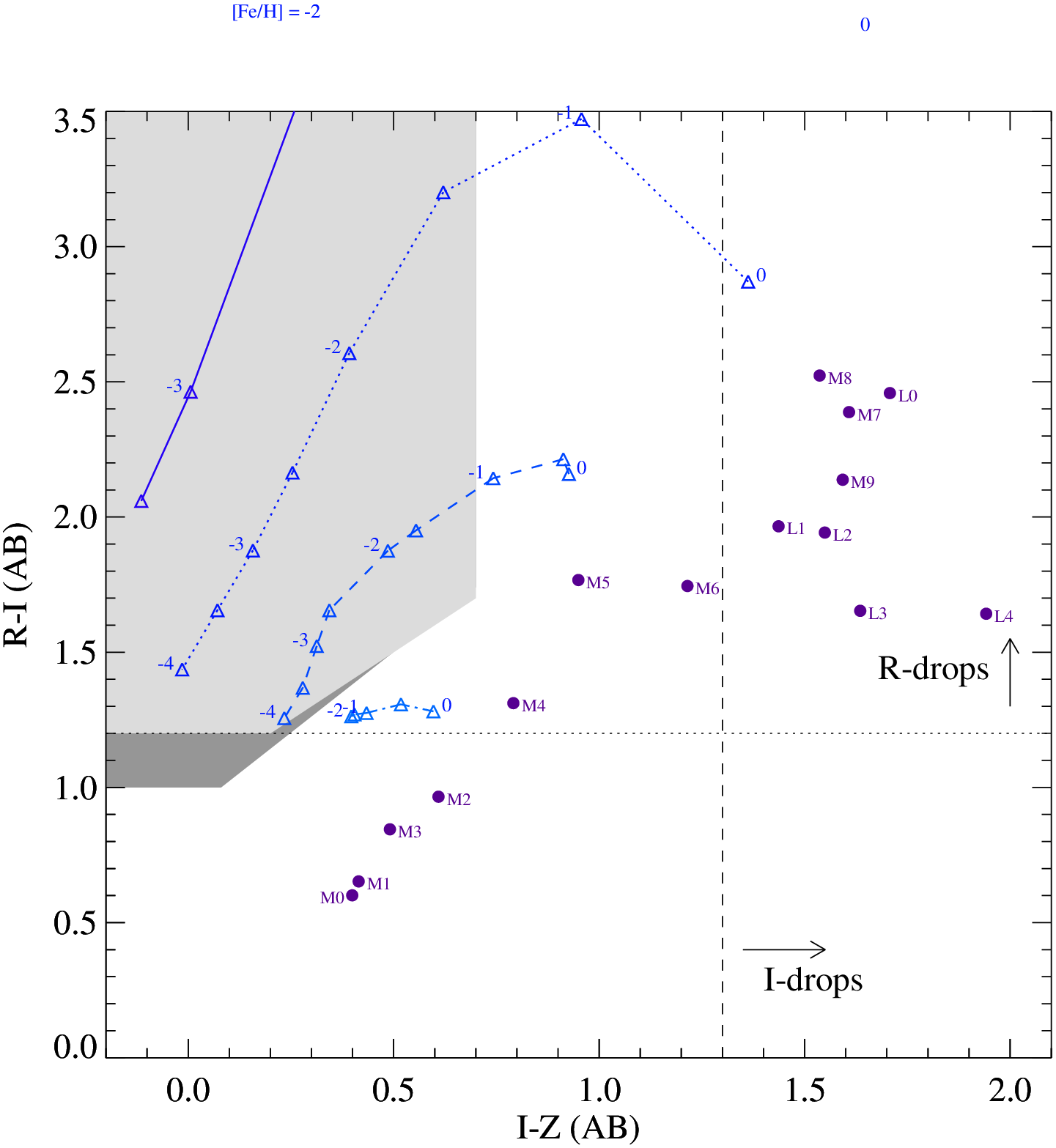}
\caption{The $R-I$ and $I-Z$ colours of model dwarf star
  atmospheres \citep[from][, triangles]{1995ApJ...445..433A},
  convolved with the Subaru/SuprimeCam filter set. Points at different
  metallicity for the same model temperature are joined by lines and
  the metallicity in terms of [Fe/H] is marked. Metallicity tracks are
  for T=2000K (solid), 2500K (dotted), 3000K (dashed) and 3500K
  (dot-dash). The colours of local cool stars from
  \citep{2002AJ....123.3409H} are shown as filled circles. The dark
  and pale shaded regions indicate the selection criteria of
  \citet{2006ApJ...653..988Y} and \citet{2004ApJ...611..660O}
  respectively.}\label{fig:metallicity}
\end{figure}

  As figures \ref{fig:biz_stars} and \ref{fig:metallicity} illustrate,
  the reliability of such a technique depends both on the depth of the
  imaging available and the metallicity of the stellar population.
  
  In figure \ref{fig:biz_stars} the spectra of M and L class stars
  \citep[constructed from a relatively bright sample in the
  Sloan Digital Sky Survey or SDSS, ][]{2002AJ....123.3409H} are convolved with the filter profiles
  of {\em HST}/ACS and those used by VLT and Subaru surveys. While the
  $B-I$ colours of cool galactic stars are less extreme than the
  technically infinite colour of high redshift galaxies in these
  bands, they can nonetheless reach very large colour decrements.  To
  reliably exclude faint stars from a survey reaching $z=5.5$ would
  require $B$-band imaging some five magnitudes deeper than the survey
  $I$-band limit.  For example, given the $i'=26.5$ limit
  of the Subaru Deep Field $Riz$ selection of
  \citet{2004ApJ...611..660O}, $B$ band imaging to $B=31.5$ would
  be required to reliably eliminate the majority of late M and early L
  class stars from the selection. The actual limiting depth of the
  Subaru Deep Field $B=27.8$ is insufficient to reliably exclude stars
  with $R$-drop colours.
   However, there is a small redshift region around $z=5$ in most
  filter sets which should be clear of stellar contamination if the
  SDSS stellar templates are representative of stars at fainter
  magnitudes. 
  
  \citet{star_paper} discuss the properties of a sample of M-stars
  selected as unresolved $v$-band dropouts ($v_{660}-i'_{775}>1.3$)
  using deep {\em HST}/ACS imaging of the Great Observatories Origins
  Deep Survey \citep[GOODS, ][]{2004ApJ...600L..93G} reaching a limit
  of $i'=25$. The optical and infrared (and, for a subsample,
  spectroscopic) properties of these unresolved sources were studied
  and found to be consistent with those of stars and inconsistent with
  those of high redshift galaxies. Stars at the faint magnitudes
  probed by high redshift surveys are likely to lie at large
  heliocentric distances and well out of the plane of the galactic
  disk. Thus a halo origin and sub-solar metallicities is likely to be
  a reasonable model for such stars.  Spectroscopic results for faint
  $M$-stars are consisted with slightly sub-solar metallicities, and a
  metallicity spread between solar and a tenth solar \citep[both
  photometrically and spectroscopically,][]{star_paper}, a range which
  can change the colours of late M stars by 0.4 magnitudes in both
  $R-I$ and $I-Z$ (figure \ref{fig:metallicity}). Importantly, at
  early M subtypes, the effect of sub-solar metallicity is to produce
  bluer $I-Z$ colours, and a unresolved sources in the GOODS fields
  were found to lie $>$0.1\,magnitudes bluewards of the stellar locus
  expected for SDSS stars with similar $v-i'$ colours.  Hence even a
  cut in $I-Z$ is not robust against faint M stars.
  
  Combining these two effects, it is difficult or impossible to
  reliably eliminate stellar contamination in $R$-drop samples given
  optical imaging alone.  The scale of this problem for ground-based
  imaging is discussed in \citet{star_paper}. 

  In ground-based surveys distinguishing stars from high redshift
  galaxies on the basis of their compact morphology is not possible.
  We quantify the effects of stellar contamination in such
  seeing-limited imaging surveys using the selection criteria of the
  Subaru Deep Field as an example. We use the $v$-drop selected GOODS
  sample of \citet{star_paper}, transforming the $bvi'z'$ photometry
  from that measured in the {\em HST}/ACS filter set to the
  Subaru/SuprimeCam filters, using the convolution of stellar
  templates from \citet{2002AJ....123.3409H} with the appropriate
  instrument response to define the transformations as a function of
  $i'-z'$ colour.  The colour-magnitude distribution measured at high
  signal-to-noise in the deep GOODS imaging is then bootstrap
  resampled to define a population of ten thousand sources and their
  magnitudes and colours perturbed by photometric errors as a function
  of magnitude calculated from the limiting depth reported by
  \citet{2004ApJ...611..660O} in each band.
  
  If the average stellar population of the two GOODS fields is
  representative of high galactic latitudes and large heliocentric
  distances more generally, the fraction of $v$-drop selected stars
  that would simultaneously satisfy the $R-I$, $I-Z$ and $B$-band
  selection criteria of \citet{2004ApJ...611..660O} is some 5\% of the
  underlying cool stellar population. Given the surface density of
  $v$-drop stars observed by \citet{star_paper}, this equates to an
  estimated contamination of 34$\pm$12 stars satisfying the $Riz$
  selection of \citet{2004ApJ...611..660O} in the SDF and SXDS fields
  (in total 1290 arcmin$^2$). The large uncertainty in the stellar
  contaminant contribution arises primarily from the field-to-field
  variation in surface density of M and early-L class stars (counts
  vary by 34\% between the GOODS-N and GOODS-S fields).
  
  This calculation is inevitably dependent on model assumptions,
  notably that the colour transformations appropriate for bright SDSS
  stars are suitable for those at faint magnitudes, and that the
  numbercounts of faint stars at $I_{AB}=25-26$ are similar to those
  at $I_{AB}=24-25$ (which show no sign of turning over). If the
  numbercounts of M-class stars drop sharply beyond 25th magnitude,
  the fraction of the underlying stellar population satisfying the
  $Riz$ criteria will drop from 5\% to 3\%.  Clearly, larger studies
  of faint cool stars in archival HST imaging is desirable to better
  constrain the behaviour of the population at these faint
  magnitudes. However, even with these caveats, we estimate that a
  substantial fraction (20-30\%) of the 106 $Riz$ sources reported by
  \citeauthor{2004ApJ...611..660O} are potentially faint stars that
  cannot be identified through ground-based optical imaging alone.

  Fortunately, the susceptibility of $R$-drop samples to stellar
  contamination is less severe in different filter sets, and will also
  depend on the existence and depth of auxiliary imaging. The
  distribution of stars between M class subtypes creates a clear
  stellar locus, decreasing in number density with increasing
  subclass/redder colours. It is possible, therefore to attempt to
  mask the small area corresponding to the stellar locus in a
  selection, either along its length (thus excluding any galaxies
  underlying it) or only where it doesn't cross the galaxy locus (thus
  including stars with identical colours to the galaxies concerned).
  Clearly in either case, the ability to perform this separation
  relies on accurate photometry and the redshift/sub-class at which
  the loci overlap.

  In the VLT/FORS2 filter set, for example, the stellar locus crosses
  the basic galaxy track at both higher redshift and later spectral
  class than in the Subaru filter set (M5$\pm$1 rather than M3$\pm$1,
  and $z\sim5.5$ rather than $z\sim5.1$).  Since there are many fewer
  late M stars than early M stars, there will be relatively few
  contaminants, and there are also fewer galaxies at high redshift
  than lower redshift leading to a reduced contribution to the total
  (galaxy+interloper) numbercounts, although the late M stars that are
  present may be hard to remove with $B$ band imaging due to their
  extreme colours.  $V$-drop samples are also less prone to stellar
  contamination (lying further from the stellar locus) and will only
  detect the most metal-poor mid-M stars, or possibly very late M and
  early L stars at the high redshift end of a sample.
  
\begin{figure*}
\includegraphics[height=0.90\columnwidth]{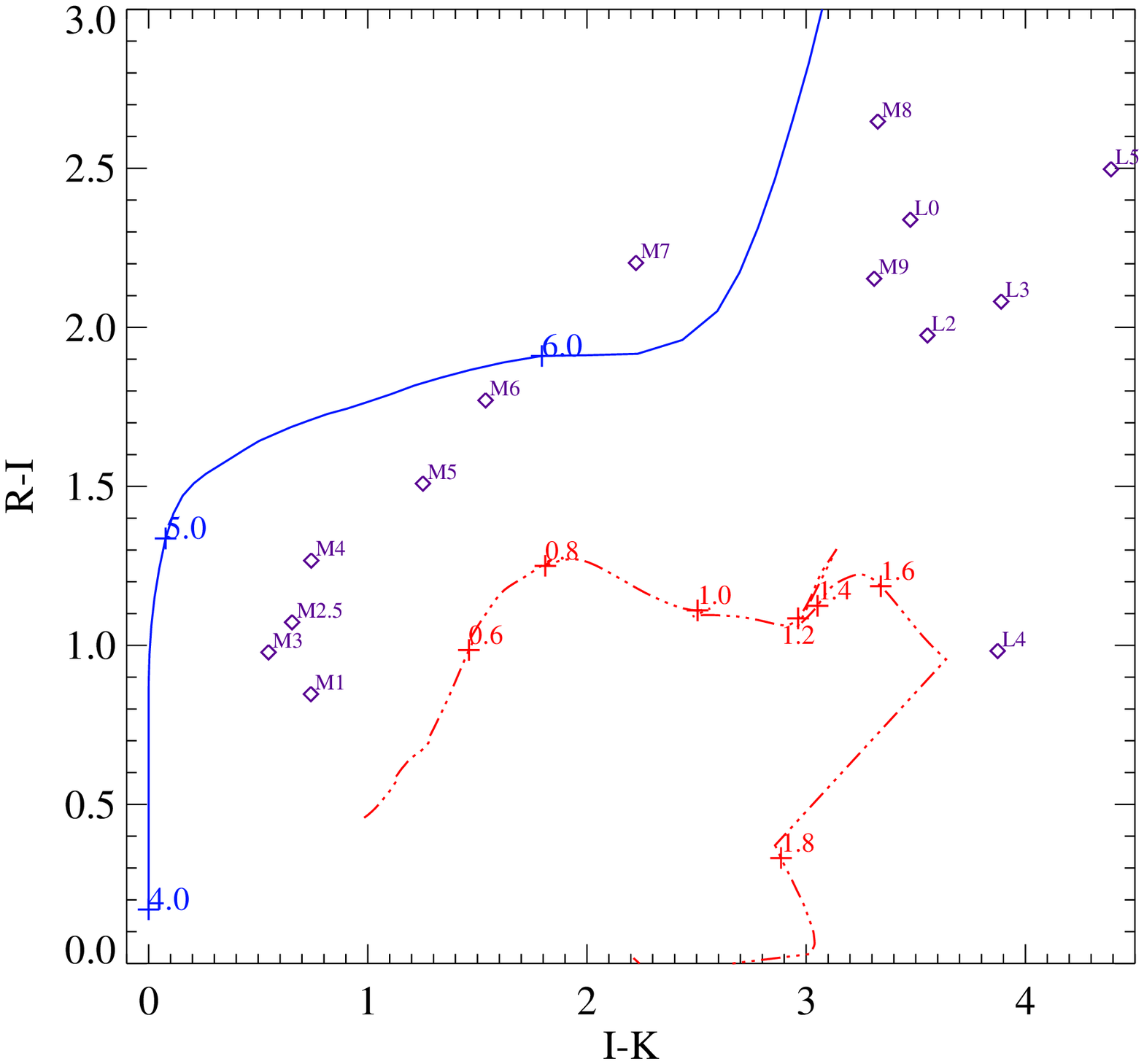}
\includegraphics[height=0.90\columnwidth]{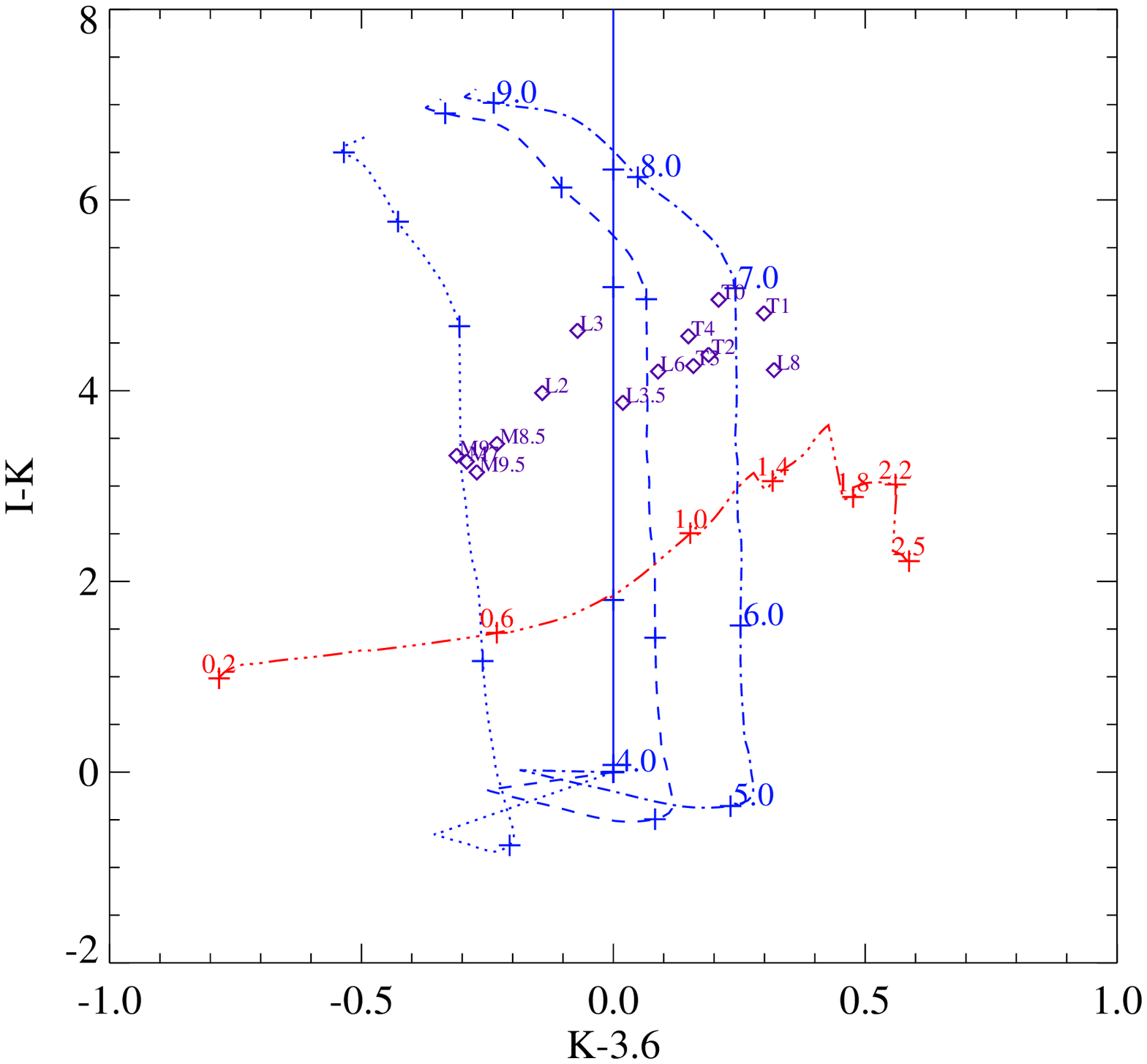}
\caption{The optical-infrared colours of cool stars (diamonds), low redshift
  elliptical galaxies \citep[dot-dot-dash][ see section
  \ref{sec:interm-galaxies}]{2005MNRAS.362..799M} and high redshift
  galaxies modelled as being flat in $f_\nu$ (solid), and with ages
  10\,Myr (dotted), 50\,Myr (dashed) and 100\,Myr (dot-dash).  Optical
  to near-infrared colours were calculated using the observed spectra
  of cool stars from \citet{2004AJ....127.3553K}, while $K-3.6\,\mu$m
  colours were derived from the results of \citet{2006ApJ...651..502P}
  Filters plotted are the VLT/FORS2 $R$ and $I$-band, the widely-used
  Mauna Kea $K_S$ band and the 3.6\,$\mu$m band of {\em Spitzer}/IRAC.
  The three populations are well-separated at $z=4-5$ but the
  distinction begins to blur around $z=6$.
  \label{fig:infrared_stars}}
\end{figure*}

  The use of a infrared colours can also help to distinguish galactic
  stars from a $z\approx5$ sample or other contaminants, although, as
  figure \ref{fig:infrared_stars} illustrates this approach becomes
  less useful with higher redshift samples. Ideally the combination of
  data in both the near-infrared and the 3.6\,$\mu$m band of the {\em
    Spitzer Space Telescope} IRAC instrument provides the cleanest
  separation of the dropout categories, since mid to late M stars can
  have comparable colours to high redshift galaxies in $I-K$. This, of
  course, presents its own difficulties for deep surveys due to the
  large point spread function and relatively shallow confusion limit
  of {\em Spitzer} when compared with optical imaging.

  \subsection{Intermediate Redshift Galaxies}
  \label{sec:interm-galaxies}
 
  The second major source of contamination in dropout samples - and
  the major contaminant from space or at faint magnitudes - is from
  low luminosity galaxies at intermediate redshifts. Such galaxies fall 
  into two categories: dusty, starforming galaxies or old, red ellipticals.
  
  Intense starbursts can lead to rapid production of large quantities
  of dust.  In the presence of dust, the rest-frame ultraviolet light
  is suppressed and re-emitted in the infrared. This results in extreme
  photometric colours that can imitate the Lyman break.  Submillimetre
  observations of distant red galaxies at $z>2$ indicate that a
  large fraction of such sources have strong starbursts, with an
  average star formation rate of 127\,M$_\odot$\,yr$^{-1}$
  \citep{2005ApJ...632L...9K}. The fraction of such sources (or their
  analogues across a range of redshifts) represented in a $V$- or
  $R$-band dropout survey is unclear, particularly since sources
  across a range of redshifts and with different reddening could
  contribute. However, spectroscopic surveys of $R$-drops have not
  reported large contamination from emission line galaxies at lower
  redshifts. Interestingly, spectroscopic follow-up to narrow band
  surveys \citep[e.g.][]{2004AJ....127..563H} has found low redshift
  line emitters, suggesting that this is a potential contaminant
  population that should be treated with caution. Infrared data should
  identify the majority of such sources, which will be bright at long
  wavelengths.

  Contamination by old, red galaxies at intermediate redshifts is more
  straightforward to quantify. Strong spectral features at longer
  wavelengths than Lyman-$\alpha$, primarily the 4000\AA\ Balmer break
  (but also strong absorption features in the blueward band), produce
  very red colours in the dropout-selection filters.  As figure
  \ref{fig:cols_filters} illustrates, such galaxies can easily satisfy a
  $V$- or $R$-drop colour selection, and with the addition of
  intrinsic variation in colour due to varied star formation histories
  can also enter an $I$-drop selection for $z\approx6$ galaxies.
  
  Although the term Extremely Red Object (or ERO) is often used as a
  short-hand for such sources \citep[e.g.][]{2004ApJ...607..704S}, the
  contaminant population for $R$-drop samples can be less extreme in
  $R-K$ or $I-K$ colour than conventional EROs \citep[defined as
  having $I-K>4$ or $R-K>5$, e.g.][]{2006MNRAS.373L..21S}.  As a
  result, $V-I$ or $R-I$ dropout samples select sources preferentially
  at $0.5<z<1.0$ and $0.6<z<1.6$ respectively (as compared to the ERO
  population which has $<z>\approx1.5$).
 
  In most galaxy surveys, this redshift range is surveyed primarily
  with the use of photometric redshifts.  While these are accurate for
  the majority of sources, they suffer from increased risk of
  catastrophic failure for atypical galaxies such as high redshift or
  very red sources, simply because the probability of obtaining a
  unique redshift solution decreases when the colours of more than one
  population become degenerate, or if rare galaxy types are not
  represented in the spectroscopically confirmed or modelled galaxy
  template set. 
  
  Some high redshift samples have exploited these large photometric
  redshift catalogues and simulated predicted contamination from
  galaxies alone yielding estimates as high as 40\% contamination for
  the $Riz$ sample of \citet{2004ApJ...611..685O} (and 26\% for their
  $Viz$ sample, in both cases not taking into account the surface
  density of stars).  Such statistical simulations of contamination
  are reasonable, although they require a detailed understanding of 
  the selection function in a given survey and thus cannot be applied
  more generally.

  \begin{figure}
\includegraphics[width=0.95\columnwidth]{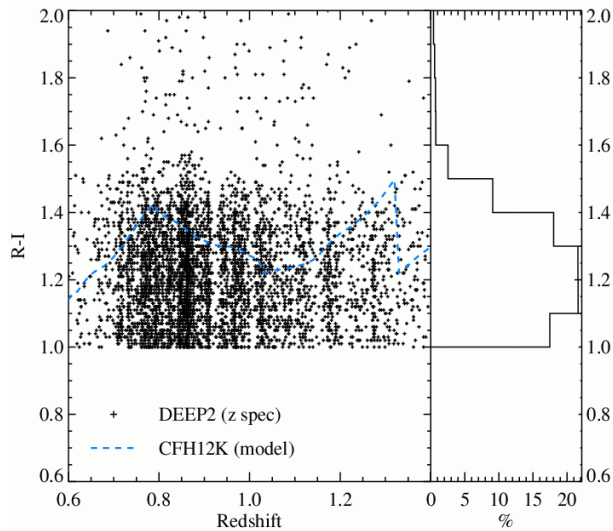}
\caption{The $R-I$ colour distribution of potential
  contaminant galaxies at intermediate redshifts. The colours of a
  mature elliptical galaxy (formed at $z_f=5$) from the population
  synthesis models of \citet{2005MNRAS.362..799M} are shown for the
  $R$ and $I$ filters of the CFH12K instrument used for the DEEP2
  survey. Points indicate the colours of spectroscopically confirmed
  $0.7<z<1.2$ DEEP2 galaxies with $R-I>1.0$.
  \label{fig:deep2_rdrops}}
\end{figure}

However, spectroscopic surveys such as DEEP2
\citep{2003SPIE.4834..161D}\footnote{http://deep.berkeley.edu/ - we
  use Data Release Two} have now characterised the population around
$z=1$ directly, measuring precise redshifts as well as $B$, $R$ and
$I$ photometry from the CFHT. As figure \ref{fig:deep2_rdrops}
illustrates, their sample includes galaxies that will easily satisfy a
cut based on $R-I$ colour.  \citet{2006ApJ...647..853W} examined the
DEEP2 galaxy luminosity function at $ 0.7<z<1.4$, dividing their
spectroscopic sample by $R-I$ colour as well as R band magnitude.
Their red galaxy sample, which is a good match for $R$-drop samples at
Subaru and the VLT\footnote{An $R-I$ cut of 1.2 in the Vega system and
  with the CFH12K instrument corresponds closely to the same colour
  cut measured in AB magnitudes in the Subaru/SuprimeCam filters},
provides a nearly-complete spectroscopic analysis of the dropout
intermediate redshift galaxies with $18.5<R_{AB}<24.1$.  Using these
data, the authors were able to fit Schecter function fits to the
luminosity function extending below the knee of the function in
redshift bins at $0.6<z<0.8$, and $0.8<z<1.0$ and to just above the
knee at $1.0<z<1.2$. In doing so, they find a typical volume density
$\phi$*$=1.35\times10^{-3}$\,gal\,Mpc$^{-3}$ at $M^*_B=-21.0$ and
$z=0.9$, with a relatively shallow faint end slope $\alpha=0.5$ with
only `modest' redshift evolution.

\begin{figure}
\includegraphics[width=0.95\columnwidth]{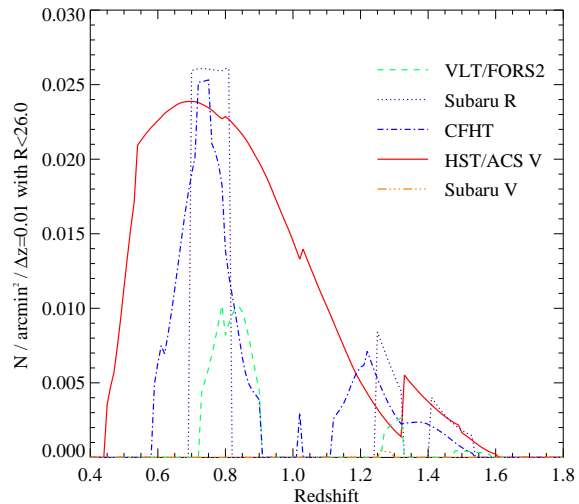}
\caption{The redshift distributions of intermediate
  redshift dropout galaxies, selected using the same colour criteria
  as given in table \ref{tab:samples} and used in figure
  \ref{fig:z_dist_rdrops}. The intermediate redshifts are modelled as
  elliptical galaxies with a formation redshift $z_f=5$ and number
  counts are generated to a limit of $I_{AB}=26$, based on the
  luminosity function derived for spectroscopically confirmed red
  galaxies at $z=0.9$ \citep[][ see text]{2006ApJ...647..853W}.  A
  typical scatter of $\pm$0.15 magnitudes is assumed, due to scatter
  in both the intrinsic colours and the photometry at these faint
  limits.
  \label{fig:z_dist_lowz}}
\end{figure}

In figure \ref{fig:z_dist_lowz} we calculate the predicted number
density of sources for a population with this luminosity function and
a \citet{2005MNRAS.362..799M} spectral energy distribution suitable
for a low redshift evolved galaxy. We use as our template the stellar
population synthesis models of \citet{2005MNRAS.362..799M} and
consider a composite stellar population forming at $z=5$
\citep[e.g.][]{2005ApJ...621..673T,2005ApJ...624L..81L} and with a
star formation rate that decays exponentially on a timescale of
0.5\,Gyr.  As figure \ref{fig:deep2_rdrops} illustrates, this provides
a reasonable fit to the properties of observed intermediate redshift
galaxies with dropout colours. We note that there is an inevitable
scatter in colour of $R$-band dropouts at intermediate redshifts due
to the variety of star formation histories. This scatter is both
bluewards and redwards of our fiducial model, but biased towards the
blue. As a result, the effect of photometric error will be to scatter
more intermediate redshift galaxies into a colour selection than out
of it.  We thus regard this model as a reasonable, if not slightly
conservative, approximation to the colour of this population.

Peaks in the redshift distribution of interlopers can be seen when the
4000\AA\ break enters the I band and also at redshifts where a
combination of emission features (such as [O II] $\lambda$ 3727\AA) in
the redward band and absorption in the blueward band (such as the
calcium and magnesium features) boosts the dropout colour.
  
  However figure \ref{fig:z_dist_lowz} does not incorporate
  constraints on the colours of these sources bluewards of the break.
  As figure \ref{fig:deep2_biz} illustrates, evolved galaxies (63\% of
  the sample) can have colours of $B-I>4$ at $0.5<z<1.2$ in all the
  filter sets discussed here.  Hence, the majority of
  contaminant galaxies can only be removed with confidence at bright
  magnitudes ($I<24$ in the Subaru Deep Field, for example), while a
  substantial population of faint contaminants is likely to remain.

Assuming that the luminosity function and $B-I$ distribution discussed
above are appropriate approximately one magnitude deeper than probed
by the DEEP2 survey, it is possible to estimate the contamination from
intermediate redshift galaxies that satisfy all the colour constraints
of the Subaru Deep Field survey (including 1\,$\sigma$ non-detection
in the $B$ band). Given these constraints, a surface density of 0.05
gal arcmin$^{-2}$, or a total $Riz$ interloper count of 64 galaxies
(in a total of 106 $Riz$ sources) would be expected in the analysis of
\citet{2004ApJ...611..660O}.  The surface densities of intermediate
redshift interlopers are comparable to those expected for $z\approx5$
galaxies \citep[e.g. 0.11 gal arcmin$^{-2}$ to $i'=26$ in the
GOODS-S,][]{2004MNRAS.347L...7B}, and broadly consistent with the 40\%
low redshift galaxy contamination estimated by
\citet{2004ApJ...611..660O} from photometric redshift catalogues and
simulations.

As figures \ref{fig:z_dist_rdrops} and \ref{fig:z_dist_lowz} make
clear, differences in filter profile can have significant effects on
the contaminant distribution and fraction from intermediate redshift
galaxies.  

As regards the V band selections, the Subaru $Viz$ selection is
cleaner than that of the {\em HST}/ACS in large part due to avoiding
the interloper galaxy tracks with its colour criteria.  However this
is not the sole reason.  It is also critical that the $V-I$ colours of
high redshift galaxies in the Subaru bands continue to increase
rapidly at $z>5$ while those measured using ACS turn over (see figure
\ref{fig:cols_filters}).  As a result there is never as much
separation between the galaxy loci at high and low redshift. It
follows that even a two-colour selection in the {\em HST}/ACS colours
cannot be as clean as that in Subaru given scatter in the low redshift
galaxy locus.

Similarly, although the R-drop selection of VLT/FORS2 is a single
colour criterion, that colour limit never includes any part of the low
redshift galaxy track, but does include all the high redshift galaxies
at $z>5$, essentially without an upper redshift limit (or rather with
one set by limiting magnitude rather than colour).  The resultant
ratio of high redshift galaxies to contaminants is high, and the
absolute number of contaminants low.

By contrast the single colour criterion of CFHT incorporates brief
redshift regions between around $z=0.8$ and $z=1.3$ for which every low
redshift galaxy (and some scattering in from outside that redshift
regime) have identical colours to high redshift galaxies and will be
selected.  Hence the number of contaminants is high and the ratio of
high z galaxies to contaminants is low.

Finally, the Subaru $Riz$ selection does theoretically exclude low
redshift galaxies, but the diagonal constraint in the colour-colour
plane is within the range of photometric scatter and variation in the
intrinsic colour of interlopers across a wide redshift range.  In
addition, the $i'-z'$ constraint applied to the Subaru selection
limits the number of $z>5$ galaxies entering the selection window, as
well as the low redshift population.  Through a combination of these
effects, the Subaru R sample has a relatively low ratio of high to low
redshift galaxies selected.

In every case, the qualitative discussion above is for our fiducial
model. We note that scatter in the colour of intermediate redshift
galaxies to the red of our model will inevitably lead to more
interloper galaxies entering dropout samples, particularly in the case
of single colour selections. We also note the importance of
near-infrared imaging where available.  As in the case of stars, the
majority of interloping galaxies can be separated from a high redshift
sample using optical-infrared colours (see figure
\ref{fig:infrared_stars}).  However, given the scatter in star
formation histories and hence infrared colours contributing to this
population, such a separation is unlikely to be clean.


\section{Interpretation of Selection Functions}
\label{sec:interpretation}

  \subsection{Optimising a Dropout selection}
  \label{sec:optimising}
  
  In sections \ref{sec:completeness} and \ref{sec:contamination} we
  discussed both completeness and contamination effects that apply to
  $z\approx5$ surveys.  In many cases the effects compete, with any
  attempt to increase the completeness of a sample also increasing its
  susceptibility to contamination by lower redshift sources.
  
  This conflict between reliability and completeness is well
  understood in radio astronomy, where it applies to the difficult
  challenge of matching radio sources with their optical counterparts
  \citep{1975AJ.....80..887C}.  Radio astronomers define a likelihood
  distribution based on the surface density of both radio galaxies and 
  faint optical sources to determine the optimum combination between
  maximising completeness and minimising the number of false matches.
  
  The challenge for high redshift samples is less well defined,
  although the discussion in section \ref{sec:contamination} above
  demonstrates that it is now possible to characterise the surface
  density of contaminant sources, at least at $z\approx5$.  Taking
  into account all of the above constraints is clearly essential when
  comparing samples collected with a disparate collection of
  instruments and selection criteria, as discussed in section
  \ref{sec:literature} below. However, it may be possible to account
  for them in the early stages of a project design.  In short, is it
  possible to optimise the filter combination and depth of a survey?
  And by what criteria should the `best' selection be judged?
  
  As figures \ref{fig:cols_rdrops} and \ref{fig:z_dist_rdrops}
  illustrate, a clean-edged redshift distribution is best attained
  using a selection colour that varies smoothly with redshift.  This
  has the added advantage of allowing a crude photometric redshift to
  be determined based on colour alone for continuum sources (although
  not for emission-line galaxies). Avoiding plateaus and
  discontinuities in the colour requires filter sets in which the $R$
  and $I$ bands neither overlap to a significant degree nor leave an
  unprobed redshift region between them.  The ideal of square-sided
  filter response curves, abutting one another in wavelength, would
  provide the smoothest variation in colour with redshift but is
  unattainable given the limitations of interference filters.
  
  Even in this ideal, no magnitude-limited sample is going to present
  a constant selection function across the $R$- or $V$-drop redshift
  range. The effects of the galaxy luminosity function (which is still
  poorly known from spectroscopically confirmed sources at $z>5$) must
  be taken into account when calculating the intrinsic properties of
  any resulting sample.

  Simultaneously minimising the filter overlap and separation has
  important consequences too for the equivalent width-dependent
  selection function of a survey.  No survey based on a simple dropout
  criterion is going to be simultaneously complete for Lyman-$\alpha$
  emitting and absorbing galaxies over a given redshift range, without
  also including many galaxies lying outside that range. The simpler
  the selection function in terms of Lyman-$\alpha$ redshift, the more
  easily this important completeness issue can be modelled.  
  
  Surveys aiming for easy photometric followup may wish to prioritise
  sources lying well away from the main high-redshift galaxy locus in
  order to secure detection of Lyman-$\alpha$. It follows that a
  spectroscopic survey in which all confirmed sources are either very
  blue in $I-Z$ or very red in $R-I$ is likely to be significantly
  incomplete of Lyman-break sources at the same magnitude limit.

  \begin{figure}
\includegraphics[width=0.95\columnwidth]{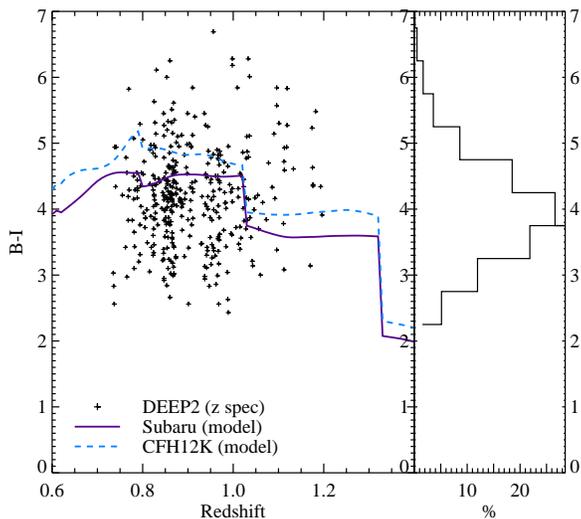}
\caption{The $B$-band depth required to eliminate
  contaminant galaxies at intermediate redshifts. The colours of a
  mature elliptical galaxy (formed at $z_f=5$) from the population
  synthesis models of \citet{2005MNRAS.362..799M} are shown for the
  $B$ and $I$ filters of the Subaru dataset and for the CFH12K used
  for the DEEP2 survey (see text). Spectroscopically confirmed
  $0.7<z<1.2$ galaxies satisfying $(R-I)_{AB}>1.4$ from the DEEP2
  survey are shown for comparison.
  \label{fig:deep2_biz}}
\end{figure}

Of course, such a survey must also account for the other half of the
completeness versus reliability dilemma.
  
  The high surface density of contaminant sources discussed in
  sections \ref{sec:cool-galactic-stars} and \ref{sec:interm-galaxies}
  highlights the importance of modelling contamination in any given
  filter set not only from intermediate redshift galaxies but also
  from extreme Galactic stars.  This constraint is particularly
  important for large area surveys observed from the ground since
  their shallower limits and large survey area leaves them highly
  vulnerable to stellar contamination.
  
  As shown in figures \ref{fig:biz_stars} and \ref{fig:deep2_biz},
  deep imaging in bands shortwards of the Lyman-limit at $z\approx5$
  can help to eliminate a large fraction, but by no means all, of the
  contaminants.  Surveys aiming to eliminate contaminants based on
  optical photometry alone must necessarily reach exceptional depths
  in the bluewards band. Even then, given the high intrinsic scatter
  in the colours of contaminant populations, the effects of the
  contaminants on any derived results must be calculated for the
  appropriate survey (as discussed in section \ref{sec:literature}).
  
  A second line of defence against contaminants may be obtained from
  the use of near-infrared imaging. Those stellar contaminants with
  the most extreme $B-I$ colours (i.e. very late M and L stars) are
  also those most easily detected in near-infrared, and thus the
  blueward and redward bands are complimentary in removing
  contaminants from the high-redshift dropout sample. Optical-infrared
  colours may assist in identifying contaminating galaxies at
  intermediate redshifts.  However, if the scatter in $B-I$ colour is
  typical of the range of star formation histories contributing to an
  intermediate redshift dropout sample, then a similar scatter might
  be expected longwards of the drop colours and hence the use of
  near-infrared filters cannot guarantee a clean sample.
  
  A $z>5$ survey intended to obtain the maximum completeness and
  minimum contamination from lower redshift sources requires imaging
  across a broad wavelength range, incorporating not only the dropout
  colour, but also depth-tuned imaging both to the blue and in the
  infrared. Given that dropout populations - both at high and low
  redshift - comprise sources with non-smooth SEDs, the colours are
  filter-dependent and their properties in any survey must be
  carefully calculated to determine the appropriate matched depths.
  Even then no Lyman-break survey will ever be complete for
  non-starforming, passively evolving galaxies at high-redshifts.
  
\begin{figure}

    \includegraphics[width=0.95\columnwidth,viewport= 0 0 623 510, clip]{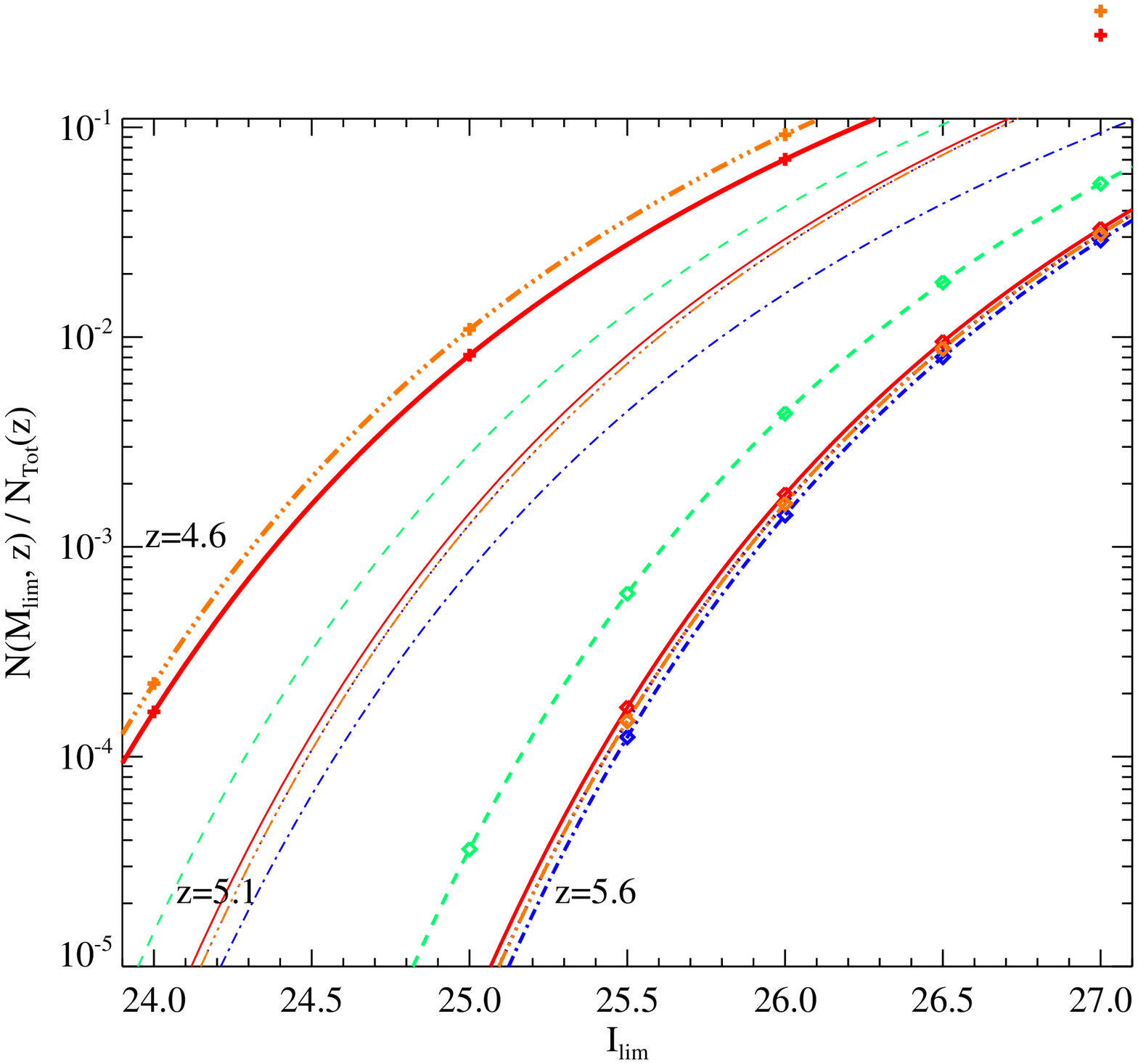}
\caption{The fraction of galaxies recovered by different survey selections
  as a function of redshift and limiting magnitude. The total number
  of galaxies at a given redshift is predicted in each case by
  integrating the luminosity function to infinity. The luminosity
  function appropriate to Lyman break galaxies at $z=3$
  \citep{1999ApJ...519....1S} is applied here at all redshifts and
  hence fractions should be considered indicative rather than
  accurate.  Changing the luminosity function alters the normalisation
  of this plot with little effect on the shape or relative response of
  different surveys (see text). Line colours and styles are as in
  figure \ref{fig:cols_rdrops}. Fractions at different redshifts are
  shown with different line thickness and symbols.
  \label{fig:completeness}}
\end{figure}

\begin{figure}
\includegraphics[width=0.95\columnwidth]{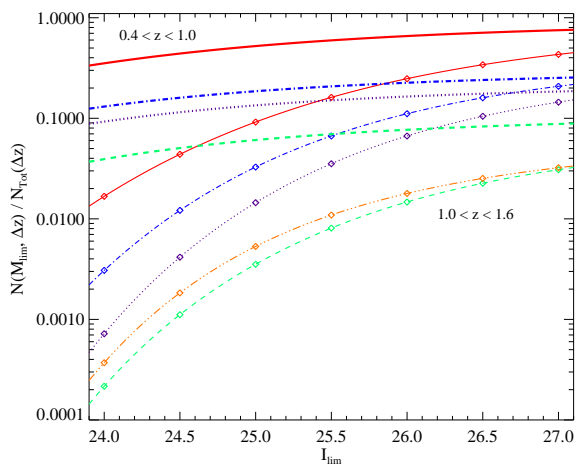}
\caption{The fraction of intermediate redshift galaxy contaminants
  satisfying the colour selection criteria shown in table
  \ref{tab:samples}, calculated from the $z=0.9$ luminosity function
  determined for spectroscopically confirmed galaxies by
  \citet{2006ApJ...647..853W}. The intermediate redshifts are modelled
  as elliptical galaxies with a formation redshift $z_f=5$ (see
  section \ref{sec:interm-galaxies}) and integrated across two
  redshift ranges: $0.4<z\le1.0$ (thick lines) and $1.0<z\le1.6$ (thin
  lines with diamonds). Line styles are as in figure
  \ref{fig:cols_rdrops}.
  \label{fig:lowz_contam}}
\end{figure}

  While noting all the points above, in figures \ref{fig:completeness}
  and \ref{fig:lowz_contam} we illustrate in basic terms the
  completeness and contamination of samples derived from the selection
  criteria in table \ref{tab:samples}. Figure \ref{fig:completeness}
  shows the fraction of all galaxies (integrated to infinite
  faintness) detectable as a function of limiting magnitude at three
  different redshifts and for each selection function. As in section
  \ref{sec:completeness}, we use the luminosity function of $z=3$
  Lyman break galaxies \citep{1999ApJ...519....1S} for reference,
  while noting that changes to the shape of the luminosity function
  make little difference to the comparative behaviour of the selection
  functions (as discussed in detail in section
  \ref{sec:filter-effect}), but have a rather larger effect on their
  normalisation.  As a result, the numerical values in figure
  \ref{fig:completeness} should be viewed as indicative rather than
  accurate.
  
\begin{figure}

    \includegraphics[width=0.95\columnwidth]{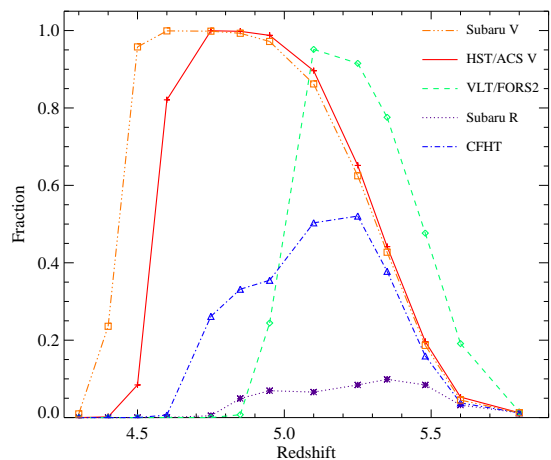}
\caption{The fraction of galaxies with $m_{1500\mathrm{\AA}}=26.0$ recovered to
  $I=26$ by different survey selections as a function of redshift. A
  typical photometric error of 0.1 magnitudes at $I=26$ was assumed in
  each case and the intrinsic colours calculated assuming a flat spectrum
 in $f_\nu$. Line colours and styles are as in figure \ref{fig:cols_rdrops}. 
  \label{fig:completeness2}}
\end{figure}

  In figure \ref{fig:completeness2} we consider a slightly different
  parameter, showing instead the fraction of galaxies with a continuum
  magnitude measured at 1500\AA\ (rest) of $m_{1500}=26.0$, recovered
  by Monte Carlo simulations to a detection limit of $I=26.0$ as
  measured in each filter set. Model galaxies were distributed in
  continuum magnitude according to the measured $z=3$ luminosity
  function and their $I$-band magnitude and $I-Z$ colour determined as
  a function of redshift and filter sets, assuming a flat rest-UV
  continuum. Colours and magnitudes were then perturbed by random
  photometric errors, assuming a typical error of 0.1\,mag at the
  selection limit of $I=26$, and the fraction of galaxies satisfying
  the selection criteria determined.  Since each $I$ band filter is
  suppressed by IGM absorption to a different degree as a function of
  redshift, the measured $I$-band value can differ from the continuum
  magnitude by several tenths of a magnitude even for a flat spectrum
  source. As a result continuum sources close to the faint selection
  limit can be lost when measured in the $I$-band. In some cases
  sources below that limit can be pulled up into the selection,
  leading to contamination of the sample by galaxies lying at $z\ge5$
  but not strictly meeting a continuum selection criterion.
  Comparison between surveys complete to any particular magnitude is
  only possible if that limit is defined by flux in a band unaffected
  by IGM absorption.
  
  Hence figures \ref{fig:completeness} and \ref{fig:completeness2}
  represent a comparison of completeness for continuum-magnitude
  limited surveys against an ideal (infinite depth) survey,
  illustrating the effect of both redshift distribution and limiting
  magnitude on the recovery of high redshift sources. In both cases,
  the relative behaviour of the different filter combinations is
  similar.  Only the two $V$-drop samples select galaxies at $z=4.6$
  since galaxies at these redshifts are too blue to be $R$-band
  dropouts.  By contrast all five selection functions are, in theory,
  sensitive to galaxies with a flat continuum at $z=5.1$ and $z=5.6$.
  At both redshifts, the VLT/FORS2 filter set and selection function
  of \citet{2007MNRAS.376.1393D}\footnote{Note:
    \citet{2003ApJ...593..630L} also used this filter set but their
    $R-I>1.5$ cut reduces completeness at $z=5.1$ while leaving it
    unaffected at $z=5.6$.}  recovers a higher fraction of the
  galaxies posited by any reasonable luminosity function than do the
  other filter sets discussed here, with the CFHT/MegaCam standard
  filter responses and the Subaru $R$-drop selection performing least
  well.
  
  In figure \ref{fig:lowz_contam} we illustrate the other side of the
  equation, using the luminosity function determined for red galaxies
  at $z=0.9$ by \citet{2006ApJ...647..853W} to predict their selection
  efficiency by any given filter and instrument combination. In each
  case only selection on $R$ or $V$, $I$ and $Z$ is assumed since, as
  demonstrated above, any associated $B$-band is usually too shallow
  to reliably eliminate a sizable fraction of these sources.
  
  Again the VLT/FORS2 filter set performs well with low sensitivity to
  contaminants in both redshift ranges $0.4<z\le1.0$ and
  $1.0<z\le1.6$. Overall, the $V$-band selection function applied at
  Subaru/SuprimeCam by \citet{2004ApJ...611..660O} suffers least
  contamination being completely insensitive to contaminants at
  $0.4<z\le1.0$ and only weakly sensitive to them at $1.0<z\le1.6$.
  Although the broad {\em HST}/ACS $v$ and $b$-bands used by the GOODS
  survey \citep{2004ApJ...600L.103G} render it more vulnerable to
  contaminants at intermediate redshifts, these are also easier to
  identify morphologically from space.

The filter combinations discussed here most vulnerable to
contamination by intermediate redshift galaxies are the $RIZ$ filter
sets commonly used by Subaru/SuprimeCam and CFHT/Megacam. These
samples are also those most susceptible to stellar contamination as
discussed in section \ref{sec:cool-galactic-stars}. In both cases, the
relatively short central wavelengths of the filters requires that
selection colours are less extreme, and therefore admit a larger
contaminant population than in other filter sets.

  With the increasing availability of sensitive large format imagers,
  combining optical filters from multiple facilities may yield
  advantages in terms of tuning the redshift selection and required
  depth in complimentary filters.

  In designing any large Lyman break galaxy survey for galaxies 
  at $z\ge5$ it is also important to note that the best indicator
  of both contamination and completeness would be a spectroscopic
  survey reaching sensitive limits and including a large fraction
  of the candidate sources.  Without such spectroscopy, the 
  properties of a survey can only be modelled rather than measured.

  \subsection{Interpretation of the Existing Literature}
  \label{sec:literature}

  Taking into account the detailed consequences of the contamination,
  selection and completeness issues above inevitably has consequences
  for the interpretation of results published and discussed in the
  literature. As case studies we consider three results that have
  attracted attention both from other observers and from theorists,
  and put them in the context of the labyrinthine selection function
  of $z>5$ Lyman break galaxies.
  
  The blue rest-UV colours of Lyman break galaxies at high redshift
  has already been the subject of speculation and interpretation.
  While Lyman break dropout galaxies are largely consistent with young
  starbursts, their rest-UV slopes have generated comment are in many
  cases too blue to fit with standard population synthesis models
  \citep[e.g.][]{2005ApJ...634..109Y,2005MNRAS.359.1184S}.  Analysing
  a sample of galaxies from the Hubble Ultra Deep Field,
  \citet{2005ApJ...634..109Y} suggested that either a top-heavy
  initial mass function (IMF) or a lower than predicted intergalactic
  medium absorption at high redshifts could explain this result. While
  most of their galaxies have not been confirmed spectroscopically, at
  least two of the galaxies described as unusually blue have
  independent spectroscopy
  \citep{2005ApJ...621..582R,2004ApJ...607..704S} and well-detected
  Lyman-alpha emission lines.  Taking \citeauthor{2005ApJ...634..109Y}
  object \#15ab as an example, the source is known to have an emission
  line with $W_0$=70\AA\ at $z=5.4$. Given this, and the effect on
  colour illustrated in figure \ref{fig:riz_ew}, its colours
  ($v-i'=2.9\pm0.3$, $i'-z'=0.24\pm0.05$) are completely consistent
  with those of a flat spectrum source. Combining such an
  interpretation with the distribution of high equivalent widths
  observed in spectroscopy of the Hubble Ultra Deep Field
  \citep{2007MNRAS.376..727S}, it is not unreasonable to assume that
  most or all of the blue sources in this field have Lyman-$\alpha$
  emission lines (with equivalent widths explainable by standard IMFs)
  pointing to a young population with sporadic bursts of intense star
  formation \citep[see also][]{2007MNRAS.tmp..294V}.

  
  Accurate measurements of the rest-frame ultraviolet luminosity
  function of $z\approx5$ are crucial to interpreting the role of the
  population in cosmological processes such as the reionisation of the
  universe and mass build-up of the largest present day galaxies
  (given assumptions for the IMF and population age).  Determination
  of the luminosity function is limited by three key parameters: the
  area of the survey, the depth in comparison to the typical
  luminosity L* and the degree to which the photometric sample
  represents the underlying population.  Despite these limitations,
  estimated $z\approx5$ luminosity functions have already been used by
  theorists to constrain both of these processes
  \citep[e.g.][]{2006astro.ph.11799M,2006MNRAS.366..705N}.
  
  Two separate luminosity functions have been derived from $z\approx5$
  photometric selections, both observed from the ground using the
  SuprimeCam instrument, but based on different imaging filters and
  selection criteria. \citet{2006ApJ...653..988Y} conducted their
  analysis on the Subaru Deep Field using the $Viz$ and $Riz$ criteria
  described in table \ref{tab:samples}.  By contrast,
  \citet{2007MNRAS.376.1557I} based their survey on a $VIZ$ sample of
  a similar size, combining the Hubble Deep Field North and a second
  non-contiguous blank field, and substituted the $I_c$ filter for the
  $i'$ used by the Subaru Deep Field team. This filter lies redwards
  of the $i'$ and hence $z\approx5$ sources show bluer $I-Z$ colours
  in the \citeauthor{2007MNRAS.376.1557I} sample than the Subaru Deep
  Field sample, while having similar $V-I$ colours\footnote{ Note: The
    colour-colour plots in \citet{2006ApJ...653..988Y} and other
    Subaru Deep Field papers appear to show a systematic offset in
    $V-I$ with respect to those of other surveys, although this offset
    is applied uniformly to data, models and selection criteria}. 
  
  In \citet{2004ApJ...611..685O}, the Subaru Deep Field team estimate
  the fraction of low redshift galaxy contaminants in their $Viz$
  sample as 26\% based on simulations of photometric redshift
  catalogues. This fraction is in good agreement with our estimate of
  the galactic contamination. Stellar contamination is not a serious
  problem for $V$-drop surveys since cool stars are generally too red
  in $I-Z$ to be confused with high redshift galaxies. Hence the
  galaxies are representative of the total sample contamination. By
  contrast, working with the same dataset,
  \citeauthor{2006ApJ...653..988Y} estimate their contamination as
  only a few percent suggesting that a degree of uncertainty exists
  concerning the properties of the sample.
  
  Furthermore, the Subaru Deep Field $Viz$ sample is supplemented by
  the the $Riz$ sample which comprises 30 per cent of the $z\approx5$
  total.  As discussed in sections \ref{sec:cool-galactic-stars} and
  \ref{sec:interm-galaxies}, contamination is likely to be a serious
  issue in the $Riz$ selection of \citet{2006ApJ...653..988Y}, with
  known contaminant populations conceivably accounting for all of the
  detected sources, and certainly accounting for $>50$\%. Overall, as
  much as half the total $z\approx5$ sample might be accounted for by
  contaminants.

  By contrast, the \citeauthor{2007MNRAS.376.1557I} survey may be
  comparatively clean. A bluewards shift in the near-vertical $V-I$,
  $I-Z$ colour track compared to the Subaru Deep Field filters has the
  effect of increasing the separation between the colours of model
  high redshift galaxies and those of most intermediate redshift
  interlopers. Nonetheless, some contaminant fraction is likely to
  remain.  The scatter in optical SEDs of intermediate redshift
  dropout galaxies (almost four magnitudes in $B-I$, as illustrated in
  figure \ref{fig:deep2_biz}) suggests that a spread in $V-I$ colours
  of approximately one magnitude would not be unreasonable. Given that
  the separation between model high and low redshift tracks is as
  small as 0.3 magnitudes at $V-I_C=1.5$ (and in the absence of
  shorter wavelength imaging), low redshift contamination is
  inevitable and will be biased towards the low redshift end of the
  sample which dominates the total number counts, particularly at
  bright magnitudes. Given the relative redshift range over which the
  two $V$-drop samples here are susceptible to low redshift
  contamination, we estimate the contamination of the
  \citeauthor{2007MNRAS.376.1557I} sample as 5-10\%.
  
  The selection functions of the two $z\approx5$ surveys, which
  theoretically probe similar redshift ranges, clearly have massive
  consequences for the contamination fraction in their surveys.  The
  distribution on those contaminants - biased to bright magnitudes in
  the \citet{2007MNRAS.376.1557I} survey and peaking in the fainter
  $Riz$ sample of \citet{2006ApJ...653..988Y} may go some way to
  explaining the differences between their derived luminosity
  functions. The function measured by the SDF team is steeper, finding
  fewer galaxies than \citeauthor{2007MNRAS.376.1557I} at bright
  magnitudes, while finding higher surface densities at faint
  magnitudes. Contaminants are unlikely to explain all the difference
  between the two luminosity functions, but certainly constitute a
  contributing factor which must be carefully accounted for as a
  function of magnitude and redshift. As a result, it is likely that
  neither the relatively shallow luminosity function slope of
  $\alpha=-1.48\pm0.3$ \citep{2007MNRAS.376.1557I} or a steeper
  $\alpha=-1.8$ to -2.3 \citep{2006ApJ...653..988Y} accurately
  describes the population at $z\approx5$.
  
  Although modelling is shaped by observation, rather than the
  reverse, comparisons between observed photometric samples and
  numerical models support the possibility of higher than supposed
  contamination in the Subaru Deep Field Sample.
  \citet{2007MNRAS.376....2K} used the large Millennium Simulation as
  the test bed for models of galaxy formation. Although these models
  provided a good fit to galaxy number counts at bright magnitudes and
  moderate redshifts, they predicted a factor of 2-8 times fewer
  galaxies at $z>4.5$ than observed in the SDF photometric sample.
  
  Since the population faint end slope is essential for calculating
  the ionising flux contribution of the population, the same
  contamination issue may explain the discrepancy between the
  ultraviolet flux density determined from the Subaru Deep Field
  \citep{2004ApJ...611..660O} and other determinations at the same
  redshift \citep{2003ApJ...593..630L}. While the latter, based in
  part on a spectroscopic survey, found evidence for a decline in the
  ultraviolet luminosity density between $z=3$ and $z=5$\footnote{Also
    seen at z=6 \citep[e.g.][]{2004MNRAS.355..374B}},
  \citeauthor{2004ApJ...611..660O} found that the luminosity density
  shows no evidence for such a decline.  Given this uncertainty,
  reionisation theorists should be wary of accepting the results of
  any one survey or selection criterion as representative of the
  starforming population at these redshifts.


 \begin{figure}
\includegraphics[width=0.95\columnwidth]{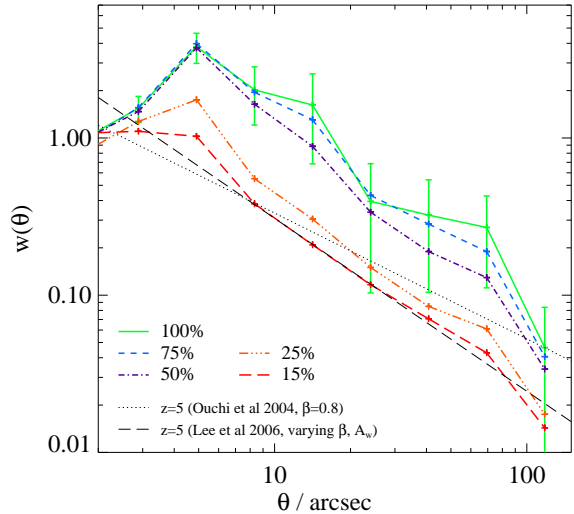}
\caption{The effect on clustering measurements of a highly clustered
  contaminant population. The clustering of spectroscopically
  confirmed $0.7<z<1.2$ galaxies satisfying $(R-I)_{AB}>1.4$ from the
  DEEP2 survey was measured, and randomly placed galaxies added such
  that the DEEP2 survey constituted between 15\% and 100\% of the total
  sample before remeasurement. The resultant population is less highly
  clustered than that at $z=1$ but still shows a strong signal, even
  with 85\% of the sample made up of randomly placed galaxies.  The
  angular clustering functions measured by \citet{2006ApJ...642...63L}
  and \citet{2004ApJ...611..685O} for $z=5$ Lyman break galaxy samples
  are shown for comparison.
  \label{fig:deep2_clustering}}
\end{figure}

  The third key result derived from Lyman break surveys, and used to
  constrain high redshift galaxy models and derive the galaxy-halo
  bias at $z\approx5$, is on the clustering of high redshift dropout galaxies.
  
  The angular two-point correlation function $w(\theta)$ describes the
  overdensity of galaxies as a function of angular scale when compared
  with a randomly generated distribution with the same
  geometry\footnote{And is defined as
    $w(\theta)$=(GG($\theta$)-2GR($\theta$)+RR($\theta$))/RR($\theta$)
    where GG is the number of galaxy-galaxy pairs at a given
    separation, RR is the equivalent number for randomly placed
    galaxies and GR is the number of galaxy-random pairs
    \citep{1993ApJ...412...64L}.}. This purely observational function
  yields an angular dependence $\beta$ and clustering amplitude $A_w$
  such that $w(\theta) = A_w \theta^{-\beta}$. These can be converted
  into a derived clustering length given assumptions about the
  redshift distribution of the sample.
  
  Measurements of the clustering at $z\approx5$ have been made based
  on dropout galaxies in the GOODS data and in the Subaru Deep Field.
  \citet{2006ApJ...642...63L} have explored the clustering of
  Lyman-Break galaxies in the GOODS fields, using the same {\em
    HST}/ACS filters but a slightly different selection function to
  that given in table \ref{tab:samples}.  This survey is one of
  relatively few to benefit from high resolution imaging data,
  allowing sources of stellar morphology to be removed before
  analysis. The selection criteria applied in the $v-i'$ vs $i'-z'$
  colour-colour plane (a slight modification of the HST/ACS criteria
  used elsewhere in this paper) should successfully remove virtually
  all lower redshift interlopers (a fact supported by visual
  inspection of the candidates). However, it is not clear that the $v$
  and $i'$ limits (just two magnitudes deeper than $z'$) are
  sufficient to support such a clean selection at the faint end of the
  sample. To cleanly select galaxies at $z=5.2$ (the upper end of
  their redshift selection) would require a $v$-band limit some three
  magnitudes deeper than $z'$.  Since it is not clear from
  \citeauthor{2006ApJ...642...63L} whether non-detections in $v$ are
  required to satisfy the $v-i'$ criterion or merely be consistent
  with it, it is possible that the sample is seriously contaminated
  with low redshift galaxies in the range $z'_{AB}=26-27$.
  
  The cost of this relatively clean selection (at least at bright
  magnitudes) is paid in allowing for relatively little 
  scatter in the high redshift galaxy colours.  Hence this result
  should be robust against any significant contamination from lower
  redshift sources but may suffer from incompleteness, particularly
  due to line emission towards the low redshift end of the selection.
  \citet{2007ApJ...659...84S} assess the
  \citeauthor{2006ApJ...642...63L} sample\footnote{and the
    \citet{2004ApJ...600L.103G} sample on which it is based} and find
  evidence for considerable incompleteness toward the low redshift end
  of their theoretical redshift selection window based on the omission
  of 12 out of 29 sources with spectroscopic redshifts in this range
  due to somewhat blue colours. The photometric selection completeness
  may well be better than this suggests since spectroscopic surveys
  are at present themselves biased towards the selection of line
  emitters in order to obtain a precise redshift.
  
  Despite this latter caveat, by excluding Galactic stars the analysis
  of clustering at $z=5$ from space-based data may well be the most
  robust measurement available, given a sensible magnitude limit,
  albeit one limited to small spatial scales and understanding of
  their true redshift sensitivity range.  The resulting measurement of
  correlation amplitude from \citeauthor{2006ApJ...642...63L} is
  slightly higher than that determined at $z\approx5$ from the SDF to
  the same magnitude limit, but consistent within the stated errors,
  and also found that the best fit required a steep angular
  dependence, $\beta=1.1$.
  
  \citet{2004ApJ...611..685O} determined $w(\theta)$ from the SDF
  $Viz$ and $Riz$ samples discussed above\footnote{We note that
    \citet{2006ApJ...637..631K} repeated this analysis, supplementing
    it with slightly deeper and simulation data and determining a
    clustering amplitude consistent with that of
    \citet{2004ApJ...611..685O}.}. While fixing the power law slope at
  $\beta=0.8$, they found evidence for a larger clustering amplitude
  at $z=5$ than observed at $z=4$ and below, but slightly lower than
  that observed by \citeauthor{2006ApJ...642...63L}. However, as
  discussed earlier, a contamination fraction of 26\% is estimated by
  the authors, and may well be much higher if the colour distribution
  of $z=1$ galaxies and cool stars are taken into account.
  
  In both \citet{2006ApJ...642...63L} and \citet{2004ApJ...611..685O},
  the authors make an adjustment for contaminant galaxies, but in
  doing so assume that such contaminants arise from photometric
  scatter and hence are distributed randomly in redshift and hence
  randomly across the sky (i.e. are unclustered).  Stellar
  contaminants satisfying $z=5$ dropout criteria are indeed unlikely
  to be significantly correlated on the sky, although
  \citet{star_paper} did find substantial variation from field to
  field. By contrast, the $z\approx1$ galaxy population discussed in
  section \ref{sec:interm-galaxies} is known to be highly clustered
  with red galaxies more biased than blue sources at the same redshift
  \citep{2007ApJ...654..115C}. As a result, the clustering signal in a
  sample may be strengthened rather than weakened by the presence of
  contaminants.
 
  In figure \ref{fig:deep2_clustering} we explore the effects of a
  small contaminant population on the clustering signal of an
  otherwise random galaxy distribution. We use as a base the
  spectroscopically confirmed $R-I>1.4$ dropout galaxies detected at
  $z=1$ by the DEEP2 survey. These were added to a randomly placed
  sample distribution of galaxies such that they comprised a varying
  fraction of the total sources, and the resultant
  clustering function was measured. As expected, the clustering amplitude
  declines as the fraction of randomly placed sources increases.
  However, as figure \ref{fig:deep2_clustering} illustrates, a
  contamination rate as low as 15\% in an otherwise unclustered sample
  can produce a measurable clustering signal. 
  
  In section \ref{sec:interm-galaxies} we calculated that
  approximately 25\% of the Subaru Deep Field $Viz$ sample is likely
  to lie at $z\approx1$.  Given the highly clustered nature of these
  contaminants, the measured angular correlation function is
  consistent with no clustering at all in the target $z=5$ population.
  Similarly the \citet{2006ApJ...642...63L} clustering result is
  consistent with no clustering at $z\approx5$ if just 15\% of their
  sample lies at $z=1$, and could have been attained within one
  standard deviation if just a few percent of their sample lies at
  $z=1$ (possible given that the numbercounts are dominated by faint
  source with non-detections in $v$).
  
  This result does not prove that the population at $z=5$ is
  unclustered (which would be very surprising given their evident
  placement in massive dark matter halos), but does highlight the fact
  that extreme care must be taken in disentangling the influence of
  highly clustered contaminant populations contributing even a few
  percent to the total.  Also interesting is that the DEEP2 data is
  too shallow to constrain the upturn in clustering seen at small
  scales by $z=5$ surveys and well detected by
  \citet{2004ApJ...611..685O} at $z=4$.  A simple visual inspection of
  high redshift candidates in {\em Hubble Space Telescope} imaging
  confirms that many of them form groups with multiple neighbours on
  the scale of a few arcseconds or less.  This small scale clustering
  is not seen at $z\approx1$ and suggests a high galaxy-halo bias.

  \subsection{Implications for Higher Redshifts}
  \label{sec:highz}
  
  In this paper we have focused on implications for galaxy surveys at
  $z\approx5$ since this is the largest and most-widely studied regime
  beyond the more easily accessible and well-known Lyman-break galaxy
  population at $z\approx3-4$. However, the same effects as those 
  discussed above are applicable to galaxy surveys targeted at higher
  redshifts.
  
  The effects of completeness-related issues are likely to become more
  severe at high redshift, due to the high background in ground-based
  near-infrared imaging.  This arises primarily from atmospheric OH
  emission in the $J$ and $H$ bands, and can vary dramatically on a
  timescale of hours.  As a result, sources at certain redshifts may
  never be observed in Lyman-$\alpha$ emission from the ground due to
  atmospheric line blanketing. The resulting broad filters, and gaps
  between filters, in the near infrared lead to less precise redshift
  discrimination, and hence potentially increased sensitivity to filter
  effects. While the common usage of the Mauna Kea filter set
  \citep{2002PASP..114..169S} is likely to mitigate this effect, where
  alternate filter definitions exist in the near-infrared they tend to
  differ dramatically in transmission profile. 
  
  In the short term, sources identified and confirmed at very high
  redshifts are likely to represent only those sources with high
  equivalent widths in Lyman-$\alpha$ and hence may not be
  representative of (or, in some cases, represented \textit{in}) the
  Lyman-break galaxy surveys that will follow.

  $I$-drop surveys, aimed at finding Lyman-break galaxies at
  $z\approx6$, are susceptible to the presence of late-M , L and T
  stars as shown in figure \ref{fig:metallicity}, while near-infrared
  dropout samples aimed at still-higher redshift may suffer
  contamination from cool, red star species such as N-type carbon
  stars \citep{2007_hawthorne,2000MNRAS.314..630T}.

Just as the redshift of Lyman-break galaxies increases as longer
wavelength filters are used for selection, so too does the redshift of
contaminant galaxies selected for their Balmer breaks. The
conventional and well studied ERO population
\citep[e.g.][]{2006MNRAS.373L..21S,2005ApJ...621...41B,2003MNRAS.346.1125G,2002A&A...381L..68C}
contributes contaminants to $z\approx6$ $I$-drop samples, while the
still-redder sources of \citet{2004ApJ...616...63Y} or the $z>2$
population discussed by \citet{2007MNRAS.376.1054D} contribute
contaminants to near-infrared dropout samples at $z>6.5$.

Although the surface density of contaminant populations decreases as
the selection moves redwards, the surface density of target high
redshift galaxies also declines sharply with increasing luminosity
distance. Hence an analogous situation to that at $z\approx5$ exists
at higher redshifts, and the issues discussed in this paper will
remain crucial to interpretations of these populations. Spectroscopy
to considerable depths will almost certainly be required to
characterise the properties of any high redshift sample. Given that
such studies form a core component of the scientific rationale for
forthcoming instruments and facilities, a firm grasp of subtle
contamination and completeness issues, and a clear reporting of the
methodology underlying any sample will be essential for some time to
come.


\section{Conclusions}
\label{sec:conclusions}

The main conclusions of this paper can be summarised as follows:

(i) The redshift distribution and number counts predicted for a given
survey is a sensitive function of filter profile, complicating
comparisons of surface density between surveys.

(ii) Moderate line emission is sufficient to move $z>5$ galaxies into
and out of selection functions, and can lead to scatter from the
continuum galaxy locus of more than a magnitude in colour for line
widths of $W_0=50$\AA. Again, the scale of this effect depends
sensitively on the filter profiles.

(iii) Results derived from ultraviolet-dropout samples apply only to a
subset of the total galaxy population. Quiescent galaxies - both young
and old - are likely to be missed by dropout samples due to faint
ultraviolet continua. Dusty galaxies at $z>5$, however, are less
likely to be omitted from samples than those at $z<4$ due to evolution
in the extinction curve.

(iv) Stellar contamination can be a serious issue for dropout
selections based on ground-based optical surveys.  Again this effect
is filter dependent, but comparison with space-based data suggests
that up to 30\% of some published samples might be accounted for by
stellar contamination alone. In these cases infrared data may help
identify contaminants.

(v) Contamination from galaxies at intermediate redshifts is again
sensitive to the filters used for colour selection. A large fraction
(63\%) of dropout galaxies at $z\approx1$ can have $B-I>4$, making
them difficult or impossible to eliminate through optical imaging
alone. The surface density of such extreme interlopers is of the same
order of magnitude as the target $z\approx5$ galaxies.

(vi) The redshift range of a survey and its susceptibility to
contamination can both be tuned by selecting different filter
combinations, ideally selecting square-sided transmission profile
filters. Deep imaging both blue and redwards of the break colours are
essential to minimise contamination.

(vii) Some contaminants, with extreme colours in all bands are only
ever likely to be identified spectroscopically. The deep imaging required
to do so photometrically is currently infeasible bluewards of the break, and
pushing the bounds of possibility in the infrared.

(viii) Care must be taken to consider selection functions and
contamination fractions when interpreting and comparing the results of
dropout surveys. Such effects may explain the discrepancies between
results at the same redshift based on different observational data.

(ix) The clustering strength seen in $z\approx5$ surveys is entirely
consistent with that expected given a small, highly-clustered
contaminant population in an otherwise uncorrelated population. While
this does not imply that $z\approx5$ galaxies are unclustered, it does
cast doubt on the reliability of current clustering measurements.

(x) The issues discussed in this paper will remain relevant at higher
redshifts, although the populations concerned and appropriate surface
densities evolve with redshift.

\section*{Acknowledgements}

ERS gratefully acknowledges support from the UK Science and
Technology Facilities Council (STFC). We thank the DEEP2 team for
making their extensive spectroscopic survey at $z\approx1$ publically
available.

%

\label{lastpage}

\end{document}